\documentclass[aip,amsmath,amssymb,reprint,
               ]{revtex4-2}

\usepackage[utf8]{inputenc}
\usepackage{bbm}

\usepackage{graphicx}
\usepackage{overpic}

\graphicspath{{Figures/}{/}}

\usepackage[english]{babel}

\usepackage{color}
\definecolor{darkblue}{rgb}{0., 0., 0.55}
\usepackage[colorlinks=true,
            urlcolor=darkblue,
            citecolor=darkblue,
            linkcolor=darkblue]{hyperref}

\newcommand{\txtd}{\textnormal{d}}

\newcommand{\expect}[1]{\ensuremath{\mathbb{E}\left[{#1}\right]}}

\newcommand{\coop}{\ensuremath{\mathrm{C}}}
\newcommand{\ncoop}{\ensuremath{n_\coop}}
\newcommand{\mcoop}{\ensuremath{n_0}}
\newcommand{\mdefect}{\ensuremath{n_1}}

\newcommand{\mcc}{\ensuremath{n_{00}}}
\newcommand{\mcd}{\ensuremath{n_{01}}}
\newcommand{\mdc}{\ensuremath{n_{10}}}
\newcommand{\mdd}{\ensuremath{n_{11}}}

\newcommand{\mccc}{\ensuremath{n_{000}}}

\newcommand{\defect}{\ensuremath{\mathrm{D}}}
\newcommand{\state}{\ensuremath{X}}
\newcommand{\actions}{\ensuremath{\mathcal{A}}}

\newcommand{\fcoop}{\ensuremath{f_0}}
\newcommand{\fdefect}{\ensuremath{f_1}}
\newcommand{\fcc}{\ensuremath{f_{00}}}
\newcommand{\fcd}{\ensuremath{f_{01}}}
\newcommand{\fdd}{\ensuremath{f_{11}}}

\newcommand{\graph}{\ensuremath{\mathcal{G}}}
\newcommand{\edges}{\ensuremath{\mathcal{E}}}
\newcommand{\triangles}{\ensuremath{\edges_3}} 
\newcommand{\nodes}{\ensuremath{\mathcal{N}}}
\newcommand{\simplex}{\ensuremath{\mathcal{S}}}
\newcommand{\degavg}{\ensuremath{k_\text{deg}}}
\newcommand{\Pxyany}{\ensuremath{P^{xy}_{\mathrm{any}}}}
\newcommand{\Pxyno}{\ensuremath{P^{xy}_{\mathrm{no}}}}

\newcommand{\Pno}[1]{\ensuremath{P^{#1}_{\mathrm{no}}}}
\newcommand{\Pany}[1]{\ensuremath{P^{#1}_{\mathrm{any}}}}
\newcommand{\Psimp}[2]{\ensuremath{P^{#1}_{\simplex_{#2}}}}
\newcommand{\excessn}[4]{\ensuremath{\mathcal{E}_{#4}^{(#1,#2)}({#3})}}
\newcommand{\simplexcess}[3]
  {\ensuremath{\mathcal{E}_{\simplex_{#2}}^{#1}({#3})}}
\newcommand{\prob}{\ensuremath{P}}

\begin{document}


\title{Stability analysis of multiplayer games on adaptive simplicial complexes}

\author{Daniela Schlager}
\affiliation{%
Technical University of Munich, Department of Mathematics,
85748 Garching bei München, Germany
}%

\author{Konstantin Clau\ss}
\affiliation{%
Technical University of Munich, Department of Mathematics,
85748 Garching bei München, Germany
}%

\author{Christian Kuehn}
\affiliation{%
Technical University of Munich, Department of Mathematics,
85748 Garching bei München, Germany
}%
\affiliation{%
Complexity Science Hub Vienna, 1070 Vienna, Austria
}

\date{\today}

\begin{abstract}
   
    We analyze the influence of multiplayer interactions
    and network adaptation on the stability of equilibrium points
    in evolutionary games.
    We consider the Snowdrift game with both two-player and three-player interactions
    on simplicial complexes.
    The state of the system and the topology of the interactions are both
    adaptive through best response strategies of nodes and
    rewiring strategies of edges, respectively.
    We derive a closed set of low-dimensional differential equations using pairwise moment closure,
    which yields an approximation of the lower moments of the system.
    We numerically confirm the validity of these moment equations.
    Moreover, we demonstrate that the stability of the fixed points
    remains unchanged for the considered adaption process. This stability result indicates that rational best response strategies in games are very difficult to destabilize, even if higher-order interactions are taken into account.
\end{abstract}

\maketitle

\begin{quotation}
    Social systems are characterized by complex networks with
    interactions of two or more agents
    (e.g., individuals, group representatives, companies).
    In game theoretic models of these systems, agents
    get a payoff depending on their chosen action and
    their neighbors actions.
    This article extends the well-established class of
    adaptive two-player games on networks, where
    the state of the system and the topology of the network
    changes over time, to three-player games on adaptive simplicial
    complexes.
    As an example the Snowdrift game with rational adaption rules
    is considered. We study the network using direct simulations as 
	well as moment closure leading to a low-dimensional differential 
	equation. We obtain a stable equilibrium state with coexistence of 
	cooperation and defection, which is remarkably robust across very large ranges of parameters.
\end{quotation}

\section{Introduction}

Complex networks are a well-established tool for the description of dynamical processes with interaction between adjacent nodes/agents. In many real world systems, however, interactions of more than two nodes take place and have an important impact on the dynamics of the system \cite{battistonNetworksPairwiseInteractions2020}, such as transmission of diseases from one infected to many others, social science~\cite{bensonHigherorderOrganizationComplex2016}, biological systems \cite{grilliHigherorderInteractionsStabilize2017}, neuroscience \cite{sizemoreCliquesCavitiesHuman2018,petriHomologicalScaffoldsBrain2014} and many more. Here, we are interested in the influence of higher-order (or polyadic) interactions in the context of game theory. So far, these multi-player interactions have been considered on fixed hypergraphs, where the structure or topology remains constant~\cite{gomez-gardenesEvolutionaryGamesDefined2011,
               penaBipartiteGraphsModels2012,
               percEvolutionaryDynamicsGroup2013}.
In contrast, adaptive networks~\cite{grossAdaptiveCoevolutionaryNetworks2008} have been successfully applied to many model systems with two-node interactions, e.g., in neural processing \cite{hernandezMultilayerAdaptiveNetworks2018}, epidemic spreading \cite{grossEpidemicDynamicsAdaptive2006}, opinion forming \cite{durrettGraphFissionEvolving2012},
and evolutionary social games
\cite{pachecoCoevolutionStrategyStructure2006,%
    kuehnEarlyWarningSigns2015,Zschaler,
    zschalerHomoclinicRouteAsymptotic2010a,
zimmermannCoevolutionDynamicalStates2004,
zimmermannCooperationSocialNetworks2005}.
The latter are often characterized by a social dilemma, in which the global payoff is not maximal, if individuals optimize their own payoff, as in the Prisoner's Dilemma or the Snowdrift game. 

Much less is known about adaptive multiplayer games on hypergraphs, where both the hypergraph and the state of the nodes evolves with time. Recently, the extension of the voter model to adaptive simplicial complexes has been investigated~\cite{horstmeyerAdaptiveVoterModel2020,MAAB} as simplicial complexes are mathematically an easier starting point in comparison to general hypergraphs. It was shown in Ref.~[\onlinecite{horstmeyerAdaptiveVoterModel2020}] that higher-order peer pressure interaction can lead to stronger fragmentation and polarization of opinions than is expected from classical two-node interactions. In fact, it is understood that higher-order coupling for static topologies can already lead to important new effects in many dynamical process, see e.g., Refs.~[\onlinecite{bickChaosGenericallyCoupled2016,%
    skardalHigherOrderInteractions2020,%
    millanExplosiveHigherOrderKuramoto2020,%
    bohleCoupledHypergraphMaps2021,%
    kuehnUniversalRouteExplosive2021}]. Hence, combining the previously mentioned directions, it becomes a natural question, how evolutionary social games behave with additional higher-order interactions and adaptive topology of the system.

Multiplayer games without adaptivity
have been studied as a model class for quite some time. For example, in the multiplayer Snowdrift game the density of cooperators drops in a replicator dynamics with the number of players and cost-to-benefit ratio~\cite{zhengCooperativeBehaviorModel2007}, there are multiplayer extensions which are equivalent to a generalized public goods game \cite{hauertSynergyDiscountingCooperation2006}, and spatial extensions have been investigated~\cite{doebeliModelsCooperationBased2005,chiongMultiagentBasedMigration2013,suiEvolutionaryDynamicsNperson2015}.
Yet, it remains unclear, how adaptivity of the topology influences the dynamics of multiplayer games, so we have to start from basic models. 

In this work, we define the adaptive simplicial Snowdrift game by combining explicit three-player interactions with the Snowdrift game on an adaptive simplicial complex. We are interested in the question, if multiplayer interactions which increase the individual payoff, such as best response and rewiring, can destabilize the steady states of the system. To study this question, we combine analytical and numerical techniques. On the analytical level, we derive low-dimensional moment equations for the lowest moments and their closures based on pairwise approximation. Numerically, we study the dynamics by direct simulation. We observe good agreement with the time evolution of the moments in both approaches, i.e., numerical solutions of the low-dimensional ordinary differential equations describing the moments match full network simulations well. Secondly, we analyze the steady states of the moment equations. It turns out that the stability properties of the main attracting steady state is robust under adding higher-order interactions beyond the classical two-player scenario. In particular, the more classical adaptive Snowdrift model with only two-player interactions behaves very similar to the higher-order variant over large robust parameter ranges. One possible interpretation of this result is that as long as agents behave in a sufficiently rational way, higher-order interactions between them do not destroy large stability of the system. Finally, we numerically evaluate how introducing some partial irrational behavior of the agents could destabilize the system. Even in this case, it is very difficult to find instabilities. 

There are two important conclusions, we can draw from these results. First, one may conjecture that many large-scale stable economic systems remain stable even if very complex higher-order interactions are considered as long as the rules the players follow are sufficiently rational. Second, although it has been shown that higher-order interactions can change dynamics in several examples for certain parameters, it should be acknowledged that there are also many cases, where the system dynamics can be very robust to adding higher-order interactions between nodes.    

The paper is structured as follows.
In Sec.~\ref{sec:model} we introduce the adaptive
simplicial Snowdrift model.
In Sec.~\ref{sec:moment_equations} we specify the
full moment equations for the system and their closure,
with a detailed derivation given in Appendices~\ref{app:derivation-moments}
and \ref{app:moment_approximations},
and present numerical results from simulation.
The stability of equilibria of the closed moment equations are analyzed in Sec.~\ref{sec:stability_analysis}.
The paper is summarized in
Sec.~\ref{sec:summary}.

\section{Adaptive simplicial Snowdrift game}
\label{sec:model}
\subsection{Multiplayer Snowdrift game}
\label{sec:multiplayer-snowdrift}
The Snowdrift game is a famous example for a game with a social dilemma
\cite{smithEvolutionTheoryGames1982,sugdenEconomicsRightsCooperation2005}.
In this game the players may or may not contribute to some task
with a total cost $c$ and a common benefit $b$.
The cost $c$ is split even among all the coorperators (strategy $\coop$),
but each player, also the defectors (strategy $\defect$), get the same benefit,
as long as the task is done.
This leads to the most commonly used
payoff matrix of the two-player Snowdrift game \cite{doebeliModelsCooperationBased2005},
\begin{equation}
    P = 
    \begin{array}{cc}
        &
       \begin{array}{cc}
           \coop \ \ & \ \ \defect
       \end{array}\\
     \begin{array}{c}
        \coop \\ \defect
    \end{array} &
    \left(
    \begin{array}{cc}
         b - \frac{c}{2} & b-c \\
         b & 0 \\
    \end{array}
    \right)
    \end{array},
\label{eq:payoff-2P-snowdrift}
\end{equation}
where the cost $c$ is split among the cooperators, while
there is a benefit $b$ if there is at least one cooperator.
We assume that the benefit exceeds the cost,
$b > c > 0$, since otherwise defection becomes
the single dominating strategy
\cite{doebeliModelsCooperationBased2005}, i.e.,
for high costs, $b < c < 2b$,
the system is equivalent to the Prisoners Dilemma.
Choosing $b > c$ implies $P(\defect,\coop) > P(\coop,\coop) > P(\coop,\defect) > P(\defect,\defect)$,
which is the defining property  of Snowdrift games \cite{doebeliModelsCooperationBased2005}.
First, this implies that there is no dominated action and
that there are two pure Nash equilibria,
$(\coop, \defect)$ and $(\defect, \coop)$,
where deviation leads to smaller payoffs.
Secondly,
both strategies can invade a population of opposite players
such that there is a mixed evolutionary stable state with 
a fraction $p_\coop = 1 - \frac{c}{2b - c}$ of cooperating players \cite{doebeliModelsCooperationBased2005}.
However, for all $p_\coop < 1$ the total payoff is smaller than in populations with only
cooperators, since
$p_\coop^2 (b-c/2) + p_\coop (1 - p_\coop) (b + b - c) + (1 - p_\coop)^2 \cdot 0 = p_\coop (2 - p_\coop)(b - c/2) \leq b - c/2$.
Therefore, the Snowdrift game is a so-called social dilemma \cite{doebeliModelsCooperationBased2005}, where maximizing individual payoffs does not
maximize the payoff for the whole society.

There are several ways to extend the Snowdrift game to a multiplayer game
\cite{doebeliModelsCooperationBased2005,%
    zhengCooperativeBehaviorModel2007,%
    chiongMultiagentBasedMigration2013,%
    suiEvolutionaryDynamicsNperson2015}.
We consider the following payoff functions for
$N$ players, which are defined for cooperating ($\coop$)
and defecting ($\defect$) players as \cite{zhengCooperativeBehaviorModel2007}

\begin{align}
P_\coop(\ncoop) &= \begin{cases}
    0 &\text{for } \ncoop = 0,\\
    b - \frac{c}{\ncoop}
                    &\text{for } \ncoop \in [1,N],
    \end{cases}\\
P_\defect(\ncoop) &=
\begin{cases}
        0 & \text{for } \ncoop = 0,\\
        b &\text{for } \ncoop \in [1,N - 1],
\end{cases}
\end{align}
where $0 \leq n_\coop \leq N$ is the number of cooperating players C.
Here the total cost $c$ of completing the task is shared equally among all cooperating players. For $N=2$ the payoff matrix in Eq.~\eqref{eq:payoff-2P-snowdrift} is recovered.

The properties of the multiplayer Snowdrift game are similar to
the  two-player game.
There are no dominated actions.
Best response against all-defecting opponents is to choose
cooperation, since the payoff increases,
$P_\coop(\ncoop=1) = b - c > 0 = P_\defect(\ncoop=0)$.
On the other hand, if there are already $l \geq 1$ cooperating opponents, 
best response is to choose defection, because
$ P_\defect(\ncoop = l) = b > b - \frac{c}{l+1} = P_\coop(\ncoop = l+1)\ \forall l \in [1,N-1]$.
Again, this implies that all pure strategy profiles, in which exactly one player cooperates, $\ncoop = 1$, are pure Nash equilibria and that there is some mixed evolutionary
stable strategy.

\subsection{Adaptive networks}

We start by defining a model class for evolutionary dynamics of the Snowdrift game on
adaptive hypergraphs.   %
A hypergraph $\graph = (\nodes, \edges)$ with up to $d$-player interactions
consist of a discrete set of nodes $\nodes$
and a set of (hyper)edges
$\edges = \bigcup_{n=2}^{d} \edges_n$ connecting the nodes.
Here $\edges_n \subset \nodes^n$ is the set of $n$-dimensional hyperedges. 
The nodes $i\in\nodes$ represent individual players and we assign a
state (or label) $\state : \nodes \rightarrow \actions$ to each node, which
for the Snowdrift game is either cooperation or defection,
$\state(i) \in \{\coop, \defect\}$.
In ordinary graphs only edges $e = (i,j)\in \edges_2$ are present
and represent possible two-player interactions between the corresponding nodes.
In hypergraphs also higher dimensional edges are allowed, e.g.,
three-node hyperedges $s = (i, j, k)$.
Thus this framework naturally allows to generalize two-player games on ordinary networks/graphs
to multiplayer games on hypergraphs.

This framework allows for two different possibilities for
adaptive dynamics.
The first is the evolution of the node state
while the set of edges remains constant.
The second is the evolution of the set of edges through creation
and deletion, while the state of nodes remains unchanged.
These two types are called dynamic \emph{on} and \emph{of} the network,
respectively, and
in adaptive networks both dynamics take place \cite{grossAdaptiveNetworks2009}.

In this work, we restrict the class of hypergraphs to simplicial complexes to keep the 
analytical complexity of the models more tractable, but our approach will be extended 
in future work.
From the viewpoint of applications, the simplicial complex restriction is applicable to systems, in which
interaction of more than two agents implies that
each subset of these agents also interacts with each other,
e.g., in friendship groups.
A simplicial complex $\mathcal{S}$ is a set of simplices with
\begin{itemize}
    \setlength{\itemsep}{0pt}
    \item[(i)] $s \in \simplex \Rightarrow \partial s \subset \simplex$, and
    \item[(ii)]  for simplices $s_1, s_2 \in \simplex$ the intersection 
    satisfies
    $s_1 \cap s_2 = \emptyset$ or $s_1 \cap s_2 \in \partial s_1, \partial s_2$,
\end{itemize}
where $\partial s$ is the set of faces of simplex $s$.
Thus, the first condition means that
every face of simplices in $\simplex$ is again in $\simplex$.
A simplicial $k$-complex is a simplicial complex $\simplex$ 
with  $\dim(\mathcal{S}) := \sup_{s \in \mathcal{S}} \dim(s) = k$,
i.e., the maximal dimension of simplices in $\simplex$ is $k$ and there is
at least one such simplex.
The first step to a generalized multiplayer Snowdrift game is
to include interactions of three players.
This means that
the adaptive dynamics is realized on a simplicial $2$-complex
consisting of $0$-dimensional nodes $\nodes$, $1$-dimensional edges $\edges_2$ and
$2$-dimensional triangles $\triangles$.
The corresponding hypergraph is then given by
$\graph = (\nodes, \edges_2\cup\triangles)$.
As before, the nodes correspond to the agents, while the edges
and triangles represent two-player and three-player interactions,
respectively.
Let us remark, that in ordinary graphs one usually calls
a set of three nodes $i, j, k \in\nodes$ a triangle if each is
connected to the others directly,
$\{(i, j), (j, k), (k, i)\}\subset \edges$.
In contrast, a simplicial triangle is a tuple $(i, j, k) \in \triangles$.
For simplicity and better distinction
we will call the elements of $\triangles$ just
\emph{simplex} in the following, and leave the term triangle for the
case of three pairwise connected nodes.
Note that within a simplicial complex each simplex $(i, j, k)\in \triangles$
implies the existence of the triangle $\{(i,j)$, $(j, k)$, and $(k, i)\}$,
since each face of $(i, j, k)$ must be contained,
but not vice versa.

The state $X(i) \in \{\coop, \defect\}$
of the nodes $i\in \nodes$ is used to label
the elements of the graph as follows.
First we identify $\coop$ with label $0$ and $\defect$ with label $1$,
such that $0$-nodes are cooperating and $1$-nodes are defecting.
The number of $\coop$-nodes in the simplicial complex is $\mcoop$
while the number of $\defect$-nodes is called $\mdefect$.
Together they satisfy $N := |\nodes| = \mcoop + \mdefect$.

Similarly, the edges are classified into $00$-, $01$-, and $11$-edges.
Again, the number of $xy$-edges is denoted $n_{xy}$
and they are counted as
$n_{xy} = \sum_{i, j=1}^N A_{ji}\delta_{X(i),x}\delta_{X(j),y}$,
where $A_{ji} = 1$ if $(j,i)\in\edges_2$ and zero else, and
$\delta_{a,b}$ is the Kronecker delta.
Under the dynamics we specify below, the total number of edges will satisfy a conservation law
$M := | \edges_2 | = \mcc + \mcd + \mdc + \mdd
= \mcc + 2\mcd + \mdd = \degavg N$,
where $\degavg$ is the average degree of nodes in the network.
The simplices are characterized by their number of defecting nodes
and we call $s = (i, j, k) \in \triangles$ a simplex of type $I$
if it contains exactly $I$ defecting nodes, i.e., 
$X(i) + X(j) + X(k) = I$. 
The set of all $I$-type simplices is denoted $\simplex_I$.
Similarly, triangles $\{(i, j), (j, k), (k, l)\} \subset \edges$
are of type $I$ if they consist of exactly $I$ defecting nodes.
Adaptive dynamics either changes
node-status and thereby the labels on the network,
or it changes the simplicial hypergraph $\graph$ by 
creation or deletion of nodes, edges or simplices.
In the following we specify a set of dynamical rules
which define the adaptive simplicial Snowdrift game.

\subsection{Adaptive simplicial Snowdrift model}

The adaptive simplicial Snowdrift model is based on the Adaptive
Simplex Voter Model \cite{MAAB,horstmeyerAdaptiveVoterModel2020}.
We assume that each player has perfect knowledge about the
currently chosen actions of the other players.
Based on the structure of the simplicial $2$-complex we consider
three different operations for the adaptive dynamics.
For all operations one specific node $i\in\nodes$
(corresponding to a player) is
selected.
This player faces either one opponent along an edge $(i,j)\in\edges_2$ or two opponents through a simplex $(i,j,k)\in\triangles$,
if this exists.
The considered strategies are \emph{rewiring}, \emph{best response on
edge}, and \emph{best response on simplex}.

\emph{Rewiring} is a dynamic of the network and
changes the simplicial complex.
The selected player $i$ removes the currently chosen edge $(i, j)$
and establishes a new edge $(i, k)$ with uniform probability to some other player $k$
with node status $\state(k) = \coop$.
Rewiring from $\defect$ to $\coop$ 
strictly increases the payoff for player $i$ since
$P(\state(i), \coop) > P(\state(i), \defect)$ for $\state(i)\in\{\coop, \defect\}$.
If the removed edge $(i, j)$ was part of a
simplex $(i, j, m)$, then this simplex is
removed as well and in order to keep the number
of simplices constant a new simplex is created
on top of one existing triangle.
Hence, this action directly influences the types of existing
edges and simplices in the simplicial hypergraph.

Best response describes a change of node states and thus
is a dynamic on the network.
In \emph{best response on edges} the selected player $i$ adjusts its own action
to the best response against the chosen opponent's action in a
two-player Snowdrift game. 
Here, best response to each action is the opposite action, see Sec.~\ref{sec:multiplayer-snowdrift}.
In particular, if the opposing action is $\coop$ ($\defect$) the state of the selected player becomes $\state(i) = \defect$ ($\state(i) = \coop$).
In\emph{ best response on simplices} the selected player $i$ adjusts its own
action to the best response against the two opponents on a simplex
$(i, j , k)$ for the three-player Snowdrift game.
This means, that if one of the opponents cooperates,
$\coop \in \{\state(j), \state(k)\}$,
the selected player
chooses to defect and the state becomes $\state(i) = \defect$.
On the other hand, if both opponents defect, $\state(j) = \state(k)  = \defect$,
the selected player chooses cooperation,
 $\state(i) = \coop$.
Thus, if the state $\state(i)$ is different from
the best response action, it changes, and otherwise
it stays constant.
If it changes, not only the label of the chosen node changes, but also the types of all edges and all simplices, which this node is part of.

Based on these three operations the evolution of the multiplayer Snowdrift game on an adaptive simplicial complex is modeled as follows.
First an edge $(i, j) \in \edges_2$ is randomly chosen in the network. 
The first node in the edge performs the following operations.
If this edge is part of at least one simplex and with probability $\rho$, best response in simplices is executed by the chosen node.
For this, the simplex is uniformly chosen from all simplices of which the edge is part of. 
Otherwise, an operation on the edge is executed by the chosen node, which is with probability $\phi$ rewiring and otherwise best response on edges. 

\section{Moment equations}
\label{sec:moment_equations}
In order to analyze the dynamics of the system we derive a set of moment
equations for the lowest moments
$\mcoop$, $\mcc$ and $\mdd$ in App.~\ref{app:derivation-moments}.
The other lowest moments follow from the conservation rule for
nodes, $N = \mcoop + \mdefect$, and for edges,
$M = \mcc + 2\mcd + \mdd$.
The resulting moment equations depend on higher order moments,
such as the distribution of simplices. 
\begin{widetext}
For the number of cooperating nodes we obtain
\begin{align}\label{equ_SD_0}
    \begin{split}
        \frac{\txtd}{\txtd t}\mcoop =
        &- \frac{\mcc}{M} \left(1 - \rho\Pany{00}\right)(1 - \phi)
        \ + \ \frac{\mdd}{M} \left(1 - \rho\Pany{11}\right)(1 - \phi) \\
        & - \frac{\mcc}{M} \Pany{00}\rho 
        \left(\Psimp{00}{0} + \Psimp{00}{1}\right)
        \ - \  \frac{\mcd}{M} \Pany{01} \rho \Psimp{01}{1}
        \ + \  \frac{\mdd}{M} \Pany{11} \rho \Psimp{11}{3},
    \end{split}
\end{align}

where $\Pany{xy}$ is the probability for an
edge of type $xy$ to be part of any simplex and
$\Psimp{xy}{j}$ is the probability that an $xy$-type edge is part
of a simplex of type $j$, conditioned it is in any simplex, see App.~\ref{app:derivation-moments}.
Even more complex are the equations for the number of $00$-edges and
$11$-edges,
\begin{align}\label{equ_SD_00}
    \begin{split}
        \frac{\txtd}{\txtd t} \mcc =
        2 \Big[ &\frac{\mcd}{M}\left(1 - \rho \Pany{01}\right) \phi
        %
        \ - \ \frac{\mcc}{M} \left(1 - \rho\Pany{00}\right)
            (1 - \phi)\left(1 + \excessn{0}{0}{0}{1}\right)
        \ + \ \frac{\mdd}{M} \left(1 -\rho\Pany{11}
        \right)(1 - \phi) \excessn{1}{1}{0}{1}\\
        %
        \phantom{=}-  
        &\frac{\mcc}{M} \Pany{00}\rho\ 
        \left(2\Psimp{00}{0} + \Psimp{00}{1}\right)
        \ - \ \frac{\mcd}{M} \Pany{01}\rho\  \Psimp{01}{1}  \\
        \phantom{=}-
        &\frac{\mcc}{M} \Pany{00}\rho\ 
        \left[\Psimp{00}{0}\simplexcess{0}{0}{0}
        + \Psimp{00}{1}\simplexcess{0}{1}{0}
        \right] 
        \ - \  \frac{\mcd}{M} \Pany{01}\rho\
        \Psimp{01}{1}\simplexcess{0}{1}{0}
        \ + \ \frac{\mdd}{M} \Pany{11}\rho\
        \Psimp{11}{3}\simplexcess{1}{3}{0}
        \Big],
    \end{split}
\end{align}
and
\begin{align} \label{equ_SD_11}
    \begin{split}
        \frac{\txtd}{\txtd t} \mdd =
        2 \Big[ &- \frac{\mdd}{M}\big(1  - \rho\Pany{11}\big) 
        \ + \  \frac{\mcc}{M}\big(1 - \rho\Pany{00}\big)
        (1-\phi)\excessn{0}{0}{1}{1}
        \ - \ 
        \frac{\mdd}{M}\big(1  - \rho\Pany{11}\big)
            (1-\phi)\excessn{1}{1}{1}{1}\\
         &+  \frac{\mcc}{M}\Pany{00}\rho\ \Psimp{00}{1} 
         \ + \ \frac{\mcd}{M}\Pany{01}\rho\ \Psimp{01}{1} 
        \ - \ 2\frac{\mdd}{M}\Pany{11}\rho\ \Psimp{11}{3}\\
        & + \frac{\mcc}{M}\Pany{00}\rho\
             \left[\Psimp{00}{0}\simplexcess{0}{0}{1} 
                    + \Psimp{00}{1}\simplexcess{0}{1}{1} \right] 
        \ + \ \frac{\mcd}{M}\Pany{01}\rho\ \Psimp{01}{1}\simplexcess{0}{1}{1}
        \ - \ 
        \frac{\mdd}{M}\Pany{11}\rho\ \Psimp{11}{3}\simplexcess{1}{3}{1}
        \Big],
    \end{split}
\end{align}

where $\excessn{x}{y}{z}{j}$ is the expected number of $z$-type neighbors
of an $x$-type node which is already connected to $j$ nodes of type $y$.
and $\simplexcess{x}{j}{z}$ is the expected number of $z$-type neighbors
of an $x$-type node which is part of a simplex of type $j$, i.e.,  which contains $j$ defecting nodes.
These contributions to the changes of moments either come from
best response or from rewiring actions;
for a detailed derivation see App.~\ref{app:derivation-moments}.

\subsection{Closed moment system}
\label{sec:moment_closure}
We consider pairwise approximation to close Eqs.~\eqref{equ_SD_0}--\eqref{equ_SD_11}.
Therefore we approximate the expected numbers of neighbors of type $z$ as
$\excessn{x}{y}{z}{j} \approx \frac{n_{zxy}}{n_{xy}} \approx \frac{n_{xy}n_{yz}}{n_{y}n_{xy}}
\approx \frac{n_{yz}}{n_{y}}$ and for nodes within simplices
we use $\simplexcess{x}{j}{z} \approx \excessn{x}{1}{z}{j}$.
A detailed derivation is given in
Appendix~\ref{app:derivation-moments},
where also approximations for $\Pany{xy}$ and $\Psimp{xy}{z}$ are specified.
Furthermore, by definition we have $\Psimp{00}{0} + \Psimp{00}{1} = 1$, since $00$-type edges
which are part of any simplex must be either within $\simplex_0$ or $\simplex_1$.
This leads to the following closed system of equations for the lowest moments,
\begin{align}\label{eq:n0_closed}
    \begin{split}
        \frac{\txtd}{\txtd t}\mcoop &=
        - \frac{\mcc}{M} \Big[ \left(1 - \rho\Pany{00}\right)(1 - \phi)
        + \rho\Pany{00} 
        \Big]
        \ + \
        \frac{\mdd}{M} \Big[ \left(1 - \rho\Pany{11}\right)(1 - \phi)
        + \Pany{11} \rho \Psimp{11}{3}
        \Big]
        %
        \ - \  \frac{\mcd}{M} \Pany{01} \rho \Psimp{01}{1}
    \end{split}\\
    \label{eq:n00_closed}
    \begin{split}
       \frac{\txtd}{\txtd t} \mcc &=
        2 \left[ \frac{\mcd}{M}\left(1 - \rho \Pany{01}\right) \phi
        \quad
        - \quad \frac{\mcc}{M} \left(1 - \rho\Pany{00}\right)
        (1 - \phi)\left(1 + \frac{\mcc}{\mcoop}\right)
        \quad + \quad \frac{\mdd}{M} \left(1 -\rho\Pany{11}
        \right)(1 - \phi) \frac{\mcd}{\mdefect} \right. \\
        %
        &\phantom{= 2 }\left. -  
        \frac{\mcc}{M} \Pany{00} \rho\ 
        \left(1 + \Psimp{00}{0} +  \frac{\mcc}{\mcoop}\right)
\quad - \quad
        \frac{\mcd}{M} \Pany{01}  \rho \
        \Psimp{01}{1}\left(1 + \frac{\mcc}{\mcoop}\right)
        \quad + \quad \frac{\mdd}{M} \Pany{11}\rho\ \Psimp{11}{3}
        \frac{\mcd}{\mdefect}
        \right],
    \end{split}\\
%
%
%
    \label{eq:n11_closed}
    \begin{split}
        \frac{\txtd}{\txtd t} \mdd &=
        2\left[ - \frac{\mdd}{M}\big(1  - \rho\Pany{11}\big) 
        \quad + \quad  \frac{\mcc}{M}\big(1 - \rho\Pany{00}\big)
        (1-\phi) \frac{\mcd}{\mcoop}
        \quad - \quad 
        \frac{\mdd}{M}\big(1  - \rho\Pany{11}\big)
        (1-\phi)\frac{\mdd}{\mdefect}\right.\\
        %
        &\phantom{= 2 \ \ } \left. +  \frac{\mcc}{M}\Pany{00}\rho \ 
        \left(1 - \Psimp{00}{0} + \frac{\mcd}{\mcoop}\right)
\quad + \quad
        \frac{\mcd}{M}\Pany{01}\rho \ \Psimp{01}{1}
        \left(1 + \frac{\mcd}{\mcoop} \right)
        \quad - \quad
        \frac{\mdd}{M}\Pany{11}\rho \ \Psimp{11}{3}
        \left(2 + \frac{\mdd}{\mdefect}\right)
        \right].
    \end{split}
\end{align}
Intuitively, each term represents one of the considered actions in the
adaptive simplicial Snowdrift game, with its corresponding implications on the
numbers of labeled nodes and edges.
However, even in their closed form, a complete analytical treatment
is not likely to work.
Instead, the expected time evolution of the lowest moments from Eqs.~\eqref{eq:n0_closed}--\eqref{eq:n11_closed} is evaluated numerically.
In the following sections we compare this to full direct simulations of the adaptive
Snowdrift game on simplicial complexes.

\end{widetext}

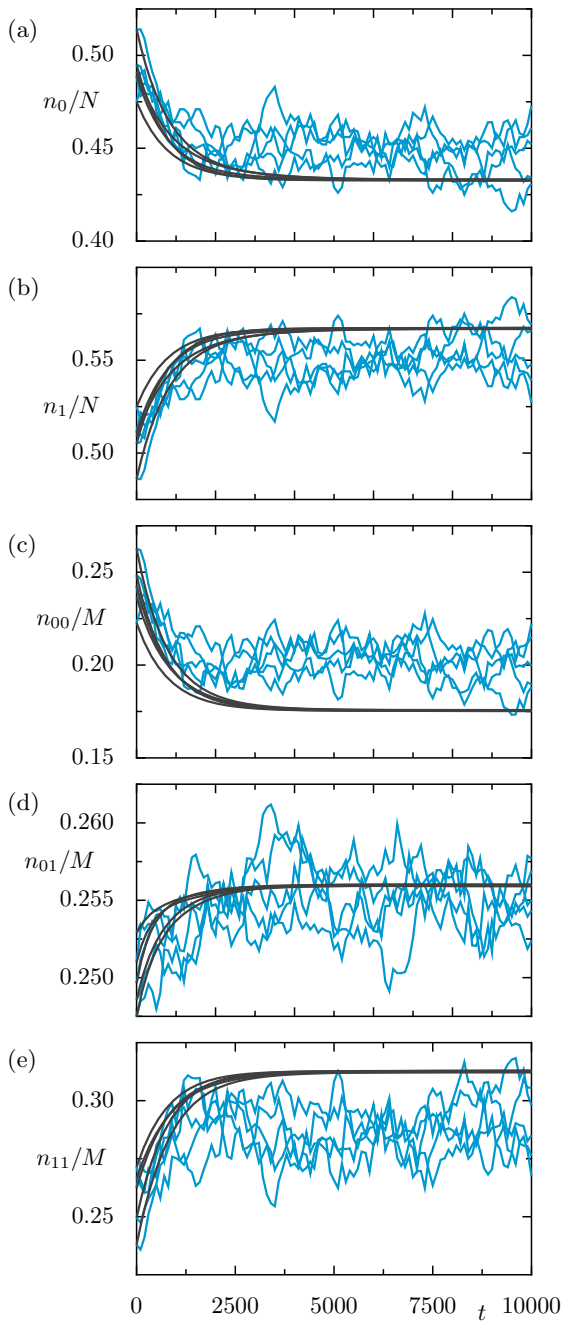
\begin{figure}[t!]
    \begin{overpic}[scale=1]
        {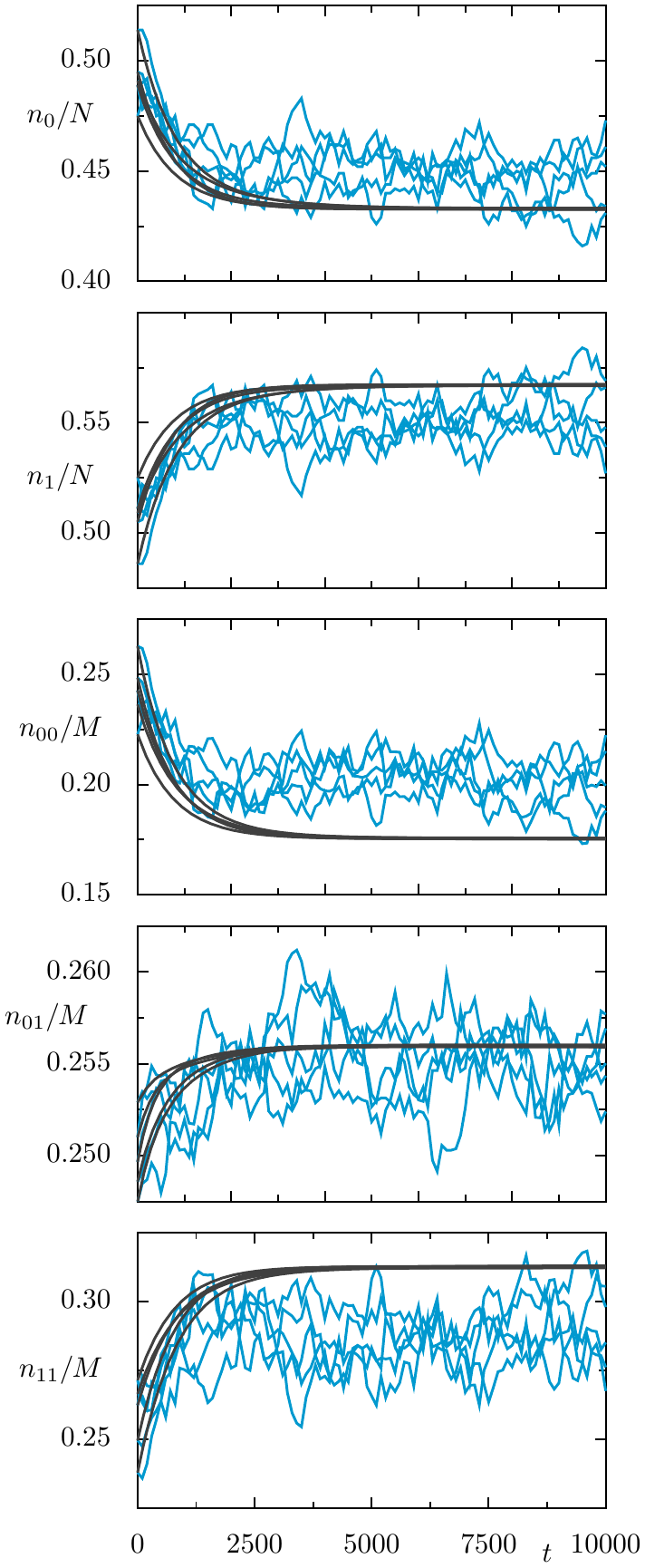}
                \put(-1, 97.5) {(a)}
                \put(-1, 78) {(b)}
                \put(-1, 58.5) {(c)}
                \put(-1, 39) {(d)}
                \put(-1, 19.5) {(e)}
    \end{overpic}
\vspace*{-.25cm}
    \caption{Numerical simulations of adaptive simplicial Snowdrift model compared to
             closed moment equations Eqs.~\eqref{eq:n0_closed} -- \eqref{eq:n11_closed}
             on random uniform networks 
             with $N =1000$ and $\mu = 19$.
             A fraction $\sigma = 0.5$ of triangles are initially considered as simplices.
            Illustrated is the time evolution of the fractions
            of
            (a) cooperators $\mcoop/N$, (b) defectors $\mdefect/N$, 
            (c) $00$-edges $\mcc / M$, (d) $01$-edges $\mcd / M$, and (e) $11$-edges $\mdd / M$
            for the simulation (blue lines) and closed moment equations (black lines).
            Other parameters are
            are $\phi = 0.5$ and $\rho= 0.5$.
            Different realizations imply a
            different number of initial simplices,
            such that there are differences in the corresponding solutions of the moment equations.
            }
    \label{fig:fig1}
\end{figure}

\subsection{Numerical results}

The closed moment equations and the approximations of subgraphs, excess degrees and probabilities used therein from Section \ref{app:moment_approximations} are compared in this section with numerical simulations on simplicial complexes.
For the simulations we
consider random Erdős–Rényi graphs
\cite{gilbertRandomGraphs1959,bollobasRandomGraphs2001} as initial networks,
with similar results for
the uniform ensemble $G(N, \tilde{M})$ and
the binomial ensemble $G(N, p)$.

Graphs in $G(N,\tilde{M})$ consist of $N$ nodes and exactly
$\tilde{M}$ undirected edges
(such that the total number of edges is
$|\edges_2| = M = 2\tilde{M}$).
Here we choose $\tilde{M}= \mu N$,
such that the average degree of each node is
$\degavg = 2\mu$.
Graphs in $G(N, p)$ consist of $N$ nodes and each of the possible
$N(N - 1)/2$ undirected edges is independently drawn with probability $p$.
Here we obtain the same total number of edges $M = \degavg N$ 
on average by choosing $p=\frac{\degavg}{N-1}$.
In order to create a simplicial complex we assign a fraction $\sigma$ of all
existing triangles as simplices in $\edges_3$.
The expected number of triangles in $G(N, p)$ is given
    by $\expect{|T|} = \binom{N}{3} p^3 $. This asymptotically
    also holds for $G(N, \tilde{M})$ with $N\rightarrow \infty$ and
    $\frac{M}{N(N-1)} = p$ such that we expect
    $\expect{|T|} = \frac{N(N-1)(N-2)}{6} \frac{\degavg^3}{(N - 1)^3}
    = \frac{1}{6}\degavg^3 \frac{N^2 - 2N}{N^2 - 2N + 1}
    \approx \frac{1}{6}\degavg^3$.

In the following, numerical results are presented
for  the uniform ensemble only, and equivalent results are
obtained for the binomial ensemble.
We choose $N=1000$ nodes,
$M = 38000$ edges (such that the average degree is $\degavg = 38$)
and $\sigma = 0.5$.
For the chosen parameters we get
approximately $9000$ triangles and $4500$ simplices
(each counted six times).%
The probability that a random edge part of at least one simplex is
approximately $P \approx 1 - \left(1 - \frac{6}{M} \right)^{\sigma\expect{|T|}}$,
which gives $P\approx 0.5$ for the chosen parameters.
Furthermore, we consider the probabilities for action on a simplex and
for rewiring of edges as $\rho = 0.5$ and $\phi = 0.5$, respectively.

The results for the lowest moments $n_x$ and $n_{xy}$ are illustrated in Fig.~\ref{fig:fig1}
for five different realizations of the initial network (blue lines).
The simulation is compared to the evolution with the closed moment equations (black lines).
We observe that within $t = 10000$ time steps the closed moment equations converge to a steady state.
The simulation shows oscillations which are in the range of the steady state of the closed moment
equations. However, there are larger deviations for the number $\mcc$ and $\mdd$ of $00$-edges and $11$-edges.

Nevertheless, the closed moment equations are a good approximation to the time evolution of
the number of nodes and edges within the adaptive simplicial Snowdrift model.
Both, simulation and closed system show in general the same trends.
and the oscillations of the number of nodes and edges in the simulation are
close to the non-trivial steady state of the closed moment equations.
We account the observed differences to the strong simplifications
made in the closure approximations, see Appendix~\ref{app:moment_approximations}.

Note that the number of simplices remains constant within each of the realizations (not shown),
because due to
$\sigma= 0.5$ there are initially enough triangles left to compensate a potential destruction of simplices by rewiring.
For different realizations, however, this number can be different due to the probabilistic number of initial
triangles and the proportional dependence of the corresponding simplices.
For the corresponding time evolution of the different number of simplex types
see Fig.~\ref{fig:fig_NumSimplex} in Appendix~\ref{app:number_simplices}.

Moreover, similar good agreement between numerical simulation and moment equations
has been checked with additional simulations for
different numbers of nodes, edges, and simplices, probabilities for actions on nodes or simplices, and initial binomial random graph $G(N, p)$.

\begin{figure}[t!]
    \begin{overpic}[scale=1]
        {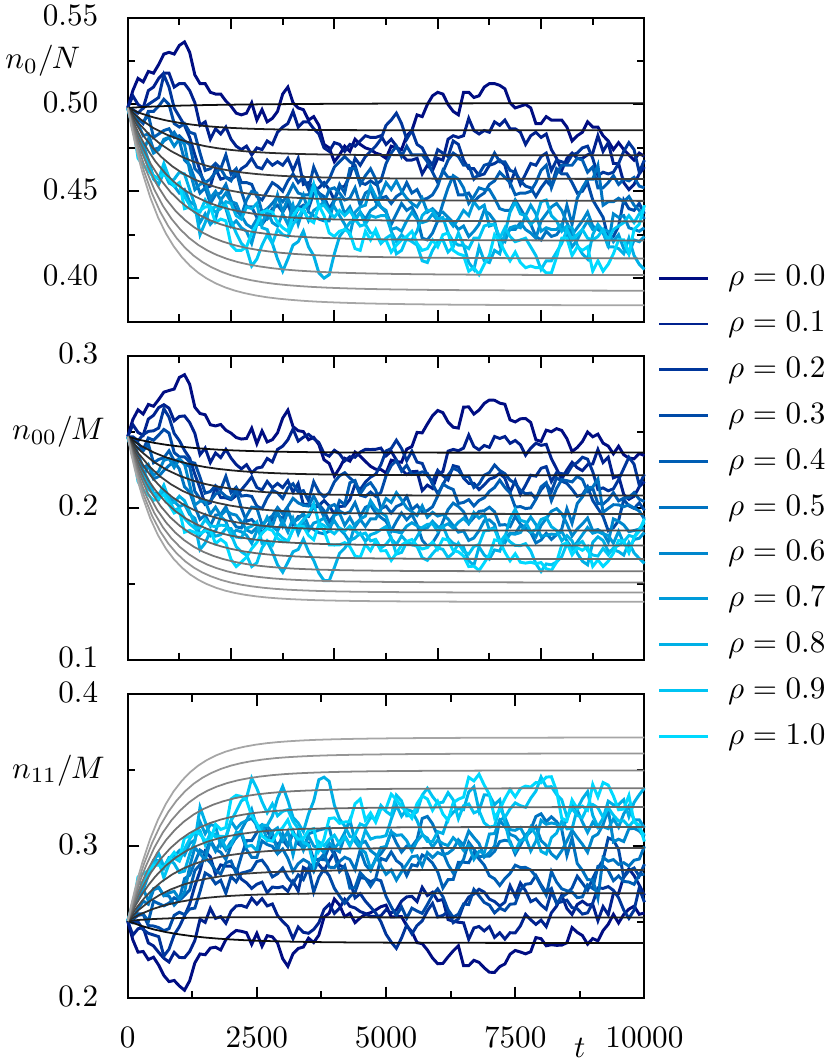}
        \put(-2, 97.5) {(a)}
        \put(-2, 65.5) {(b)}
        \put(-2, 34.5) {(c)}
    \end{overpic}
    \caption{%
        Dependence of adaptive simplicial Snowdrift model on the probability 
        of choosing an action on the simplex $\rho$.
        Shown are numerical simulations (colored lines) compared to
        closed moment equations Eqs.~\eqref{eq:n0_closed} -- \eqref{eq:n11_closed}
        (gray lines)
        with same parameters as in Fig.~\ref{fig:fig1} and with $\rho$ as specified.
        Illustrated is the time evolution of
        the fractions of
        (a) cooperators $\mcoop/N$,
        (b) $00$-edges $\mcc / M$, and
        (c) $11$-edges $\mdd / M$.
    }
    \label{fig:fig2}
\end{figure}

In Fig.~\ref{fig:fig2} we illustrate how the adaptive simplicial Snowdrift model
depends on the probability $\rho$ to choose an action on simplices.
For this, the time-dependence of the lowest moments $\mcoop$, $\mcc$ and $\mdd$
is illustrated for different values of $\rho$ as specified
in the figure (blue colored lines).
This is compared to the corresponding solution of the closed moment equations (gray colored lines).
We observe that for all $\rho$ the lowest moments quickly approach an equilibrium, around which there are fluctuations in the simulations.
The observed steady states for the moment equations
     suggest a monotone dependence on the parameter $\rho$.
      For example, the fraction of cooperators in the equilibrium
      $\fcoop = \mcoop / N$
      decreases from $\fcoop^\ast = 1/2$ (with zero multiplayer interactions) to approximately
      $\fcoop^\ast = 1/3$ (with many multiplayer interactions).
This means that increasing the influence of these multiplayer interactions by
increasing $\rho$ does not lead to a change in stability or a bifurcation.
In contrast, the observed non-trivial
steady states seem to be stable for the simplicial
Snowdrift game, even when certain irrational perturbations are considered,
as discussed in the next section.


\section{Stability analysis}\label{sec:stability_analysis}

The closed moment equations which describe the evolution of the adaptive simplicial Snowdrift game
cannot be examined completely analytically. 
However, it is possible to analyze steady states and their local stability for the different operations
rewiring, best response on edge and best response in simplices, separately. 
For this purpose we consider the fractions of labeled nodes and edges
$\fcoop=\frac{\mcoop}{N}$, $\fdefect=\frac{\mdefect}{N}=1-\fcoop$, $\fcc=\frac{\mcc}{M}$, $\fdd=\frac{\mdd}{M}$, and $\fcd=\frac{\mcd}{M}=\frac{1}{2}(1-\fcc-\fdd)$,
and rewrite the closed moment equations accordingly.

\subsection{Rewiring}
The dynamics of rewiring can be extracted from the closed moment equations (\ref{eq:n0_closed})-(\ref{eq:n11_closed}) by choosing $\phi=1, ~\rho=0$ or $\phi=1,~ S=0$.
Therefore, the corresponding dynamics of the adaptive network in relative variables is
given by
\begin{align}\label{equ_SD_rew}
    \begin{split}
        \frac{\txtd}{\txtd t} \fcoop &= 0, \\
        \frac{\txtd}{\txtd t} \fcc &= \frac{1 - \fcc - \fdd}{M},\\
        \frac{\txtd}{\txtd t} \fdd &= -2 \frac{\fdd}{M} .
    \end{split}
\end{align}
This system has a line of steady states
$\left\lbrace (\fcoop^*,\fcc^*,\fdd^*)=(\fcoop^*,1,0) ~| ~\fcoop^* \in (0,1] \right\rbrace $.
Here $\fcoop^*=0$ is excluded because without $0$-nodes there are no
$00$-edges which contradicts $\fcc^*=1$ in the steady state.
Note that in the derivation of the moment equations, App.~\ref{app:derivation-moments},
    it is assumed, that there always are enough $0$-nodes, to which the selected edge can be rewired.
    In networks with a small fraction of $0$-nodes but with high node-degrees $\mu$
    this assumption will be violated and Eq.~\eqref{equ_SD_rew} cannot be applied.

The second and third equation in Eqs.~\eqref{equ_SD_rew} are independent of $\fcoop$,
which allows to reduce the system to the $\fcc$ and $\fdd$ variables.
The Jacobian of the right hand side of the reduced system
is given by
\begin{equation}
    J(\fcc, \fdd) = \frac{1}{M}\left(
    \begin{array}{cc}
        -1 & - 1 \\ \phantom{+}0 & -2
    \end{array}
    \right) 
\end{equation}
which has two negative real eigenvalues
$\lambda_1 = -\frac{1}{M}$ and $\lambda_2 = -\frac{2}{M}$.
Therefore, the steady state is locally asymptotically stable in this reduced system.
Since the fraction of cooperators $\fcoop$ is always constant in the system,
the line of steady states defined above is stable in the system with only rewiring.

\subsection{Best response on edges}
For the parameter choices $\phi=0, ~\rho=0$ or $\phi=0, ~S=0$ in the closed moment equations \eqref{eq:n0_closed}--\eqref{eq:n11_closed} only best response on edges occurs in the adaptive network.
In terms of $\fcoop,~\fcc,~\fdd \in [0,1]$ this is given by
\begin{align}
        \frac{\txtd}{\txtd t} \fcoop =& -\frac{\fcc}{N} +\frac{\fdd}{N},\nonumber \\
        \frac{\txtd}{\txtd t} \fcc =& - 2\frac{\fcc}{M} - 2\frac{\fcc^2}{N\fcoop} +
\frac{\fdd (1 - \fcc - \fdd) }{N(1 - \fcoop)},
\label{equ_SD_bre}
\\
        \frac{\txtd}{\txtd t} \fdd=& - 2\frac{\fdd}{M} +
             \frac{\fcc (1 - \fcc - \fdd)}{N\fcoop} -\frac{2\fdd^2}{N(1 - \fcoop)},
             \nonumber
\end{align}
For $N, M \neq 0$ this system has a line of trivial steady states $\left\lbrace (\fcoop^*,\fcc^*,\fdd^*)=(\fcoop^*, 0, 0)~|~\fcoop^* \in (0,1)\right\rbrace $.
The cases $\fcoop^*=0$ and $\fcoop^*=1$ are excluded since this contradicts
$\fcc = 0$ and $\fdd = 0$.
Note that depending on $\fcoop^\ast$ there might not be enough
possible $01$-edges in simulations on some finite network.
The Jacobian in one of the steady state points $(\fcoop^*,\fcc^*,\fdd^*)=(\fcoop^*, 0, 0)$
is given by
\begin{equation}
    J(\fcoop^\ast) = \left(
    \begin{array}{ccc}
        0 & -\frac{1}{N} & \frac{1}{N} \\
        0 & -\frac{2}{M} & \frac{1}{N(1 - \fcoop^\ast)} \\
        0 & \frac{1}{N\fcoop^\ast}  & -\frac{2}{M}
    \end{array}
    \right) 
\end{equation}
with eigenvalues
$\lambda_0 = 0$ and $\lambda_{2,3} = - \frac{2}{M} \pm \frac{1}{N \sqrt{\fcoop^\ast (1 - \fcoop^\ast)}}$.
The second eigenvalue becomes positive for all $\fcoop^\ast$ if $\frac{M}{N} > \sqrt{2}$,
i.e., if the average degree $\mu > 1/\sqrt{2}$. This is satisfied for realistic networks.
The Hartman-Grobman theorem implies that under this condition each of the trivial
steady states is unstable.

Furthermore, the system Eq.~\eqref{equ_SD_bre} has a non-trivial steady state
$(\fcoop^*,\fcc^*,\fdd^*) = \big(\frac{1}{2}, \frac{1}{4}[1 - \frac{N}{M}], \frac{1}{4}[1 - \frac{N}{M}]\big)$.
Therefore, if the number of edges $M$ is sufficiently high relative to the number of nodes $N$,
this steady state corresponds to the state in which all node labels and edges are uniformly distributed,
i.e., $(\fcoop^*,\fcc^*,\fdd^*)\rightarrow\big(\frac{1}{2},\frac{1}{4},\frac{1}{4}\big)$ for $N/M \rightarrow 0$.
Calculating the eigenvalues and applying the Hartman-Grobman theorem reveals
that this steady state
is asymptotically stable for $\mu = \frac{M}{N} > 1 $, which is fulfilled for all
reasonable networks.
For large $M$ this can be seen as the realization of the only mixed Nash equilibrium of
the 2-player Snowdrift game globally in the network,
where both players select either of the two actions $\coop$ and $\defect$ with the same
probability $1/2$.

\subsection{Best response in simplices}
Choosing $\rho=1$ and a large number $S$ of simplices implies that
every edge is part of at least one simplex in Eqs.~\eqref{eq:n0_closed}-\eqref{eq:n11_closed}.
Thus, only best response in simplices is executed in the adaptive network.
In particular, here we assume that $\Pany{xy} = 1$ for all $xy$-edges and apply pair
approximation for the probabilities of $xy$-edges
to be part of specific simplex types $\Psimp{xy}{i}$, see Appendix~\ref{app:approx_prob}.
This leads to the following system in terms of $\fcoop,~\fcc,~\fdd \in [0,1]$,
\begin{widetext}
\begin{align}\label{equ_SD_brs}
    \begin{split}
        \frac{\txtd}{\txtd t} \fcoop =& -\frac{\fcc}{N}
        - \frac{\fcd}{N}\frac{\fcc \fdefect}{\fcc \fdefect+\fdd \fcoop}
        +\frac{\fdd}{N}\frac{\fdd^2 \fcoop}{3 \fcd^2 \fdefect+\fdd^2\fcoop},\\
        \frac{\txtd}{\txtd t} \fcc =
        &- 2 \frac{\fcc}{M}\Bigg(\frac{\fcc^2 \fdefect}{\fcc^2\fdefect+ 3\fcd^2\fcoop}+1\Bigg)
        - \frac{2\fcc^2}{N\fcoop}
        -2 \left(\frac{\fcd}{M} + \frac{\fcc \fcd}{N\fcoop}\right)
        \frac{\fcc \fdefect}{\fcc \fdefect+\fdd \fcoop}
        + 2 \frac{\fcd \fdd}{N \fdefect}\frac{\fdd^2\fcoop}{3\fcd^2 \fdefect+\fdd^2 \fcoop},\\
        %
        \frac{\txtd}{\txtd t} \fdd = 
        &+ 2\frac{\fcc}{M}\frac{3\fcd^2\fcoop}{\fcc^2\fdefect +  3\fcd^2\fcoop}
        + 2\frac{\fcc \fcd}{N\fcoop}
        +2\left( \frac{\fcd}{M}  + \frac{\fcd^2}{N \fcoop} \right) \frac{\fcc \fdefect}{\fcc \fdefect +\fdd \fcoop} 
        - 2\left( \frac{2\fdd}{M} + \frac{\fdd^2}{N\fdefect}\right)
        \frac{\fdd^2\fcoop}{3\fcd^2\fdefect+\fdd^2\fcoop}.
    \end{split}
\end{align}
\end{widetext}
Again we find a line of trivial steady states for best response in simplices,
$\{(\fcoop^*,\fcc^*,\fdd^*)=(\fcoop^*, 0, 0) ~|~ \fcoop^* \in (0,1)\}$,
which are the same as for best response on edges.
Further steady states can only be computed implicitly by the usage of mathematical software.

In order to analyze the dynamics of best response in simplices
besides the trivial steady states we assume in the following that
$\Psimp{xy}{i}=\frac{1}{2}$ for all possible edge and simplex types.
This is equivalent to the assumption that
the simplex types are uniformly distributed
within all simplices, i.e., their total amount is the same
$|\simplex_0| = |\simplex_1| = |\simplex_2| = |\simplex_3|$.
The resulting system in terms of $\fcoop,~\fcc,~ \fdd \in (0,1)$ is given by
\begin{align}\label{equ_SD_brs_half}
    \begin{split}
        \frac{\txtd}{\txtd t} \fcoop =& -\frac{\fcc}{N}
        -\frac{1}{2}\frac{\fcd}{N}
        + \frac{1}{2} \frac{\fdd}{N},\\
        \frac{\txtd}{\txtd t} \fcc =
        &- 3\frac{\fcc}{M} - \frac{2\fcc^2}{N\fcoop}
        - \frac{\fcd}{M}  - \frac{\fcc \fcd}{N\fcoop}
        + \frac{\fcd \fdd}{N\fdefect},\\
        \frac{\txtd}{\txtd t} \fdd = 
        &\frac{\fcc}{M} + 2\frac{\fcc \fcd}{N\fcoop}
        + \frac{\fcd}{M} +  \frac{\fcd^2}{N\fcoop}
        - 2\frac{\fdd}{M}   -\frac{\fdd^2}{N\fdefect}.
    \end{split}
\end{align}
For this simplified system we find the nontrivial steady state
$(\fcoop^*,\fcc^*,\fdd^*) = 
\Big(\frac{1}{3},
\frac{2  \mu - 2 + \sqrt{4\mu^2 + 2\mu + 4}}{9 \mu} - \frac{1}{3},
\frac{2  \mu - 2 + \sqrt{4\mu^2 + 2\mu + 4}}{9 \mu}\Big)$
with $\mu=\frac{M}{N}$.
Again 
$\fcoop^*=\frac{1}{3}$ corresponds to the fraction of cooperators $\coop$ in the Nash Equilibrium
of a 3-player Snowdrift game,
which is consistent with the assumption of initial uniform distribution of labels, edges and simplices.

\begin{figure}[t!]
    \begin{overpic}[scale=1]
        {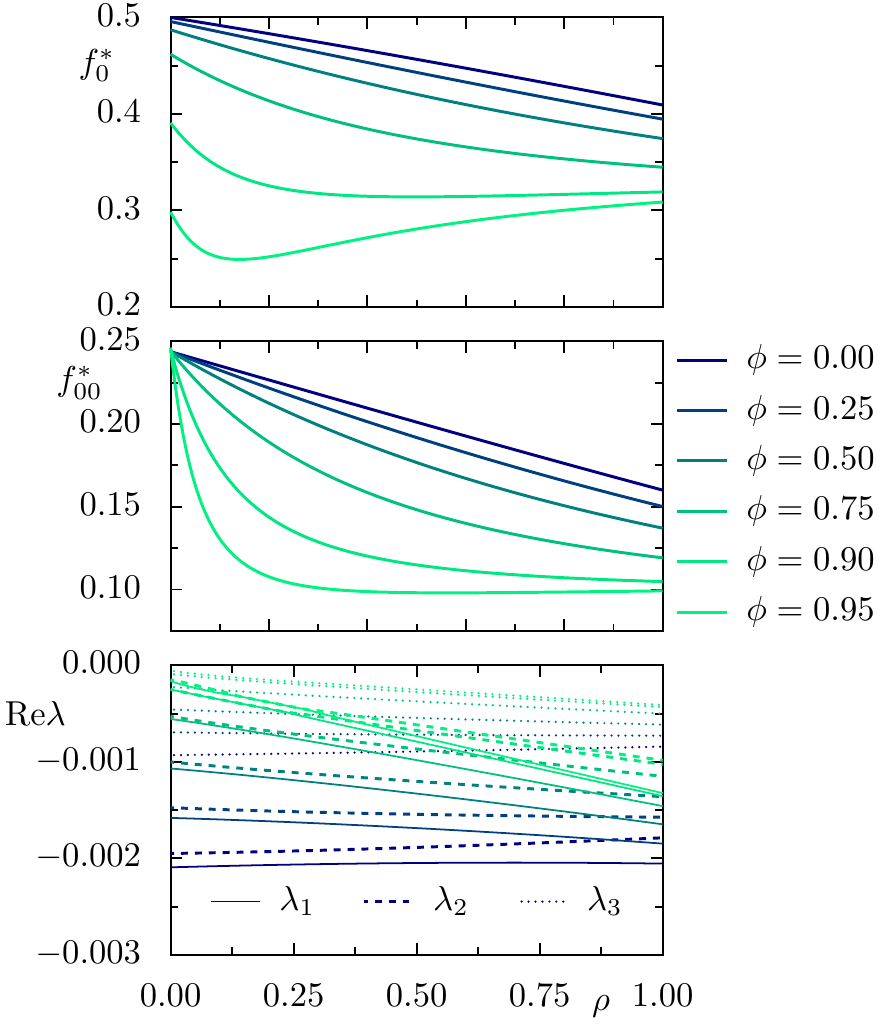}
        \put(-0, 97.5) {(a)}
        \put(-0, 65.5) {(b)}
        \put(-0, 34.5) {(c)}
    \end{overpic}
    \caption{%
        Stability analysis with MatCont.
        Shown are (a) the fraction of cooperators $\fcoop^\ast$ and (b) the fraction
        of $00$-edges $\fcc^\ast$ as a function of $\rho$
        in the equilibrium of the moment system with
        $N = 1000$, $\mu = 20$, $\sigma = 0.5$ for different values of $\phi$.
        In (c) the corresponding, numerically obtained real parts of
        the eigenvalues $\lambda_i$ are illustrated,
        which are smaller than zero for all $\rho$.
    }
    \label{fig:fig_matcont}
\end{figure}
\subsection{Numerical analysis with MatCont}

We perform a numerical stability and bifurcation
analysis of the adaptive simplicial Snowdrift model with the help of
the software tool MatCont \cite{dhoogeNewFeaturesSoftware2008} as described in Appendix~\ref{app:matcont}.
The numerical analysis confirms the analytical results of the
equations for rewiring, best response on edges and best response in simplices separately.
Furthermore, the analysis in MatCont indicates that the closed moment equations, Eqs.~\eqref{eq:n0_closed}--\eqref{eq:n11_closed},
of the full Adaptive Simplex Snowdrift Model
have a non-trivial steady state which is asymptotically stable for
all reasonable parameter choices
$\phi \in [0,1]$, $\rho \in [0,1]$, $N > 3$, $\mu > 1$ and
$\sigma \in [0, 1]$.

An exemplary parameter diagram is shown in Fig.~\ref{fig:fig_matcont} as a function of
$\rho$ and for different parameters $\phi$. The fractions of cooperators and $00$-edges in
the equilibrium are shown in panels (a) and (b), while the real parts of the
corresponding eigenvalues are shown in (c). Apparently, all eigenvalues are below zero,
such that the steady states are stable for all parameters. Yet, we observe that the eigenvalues
are relatively close to zero so that the stability is relatively weak, which does allow after 
a perturbation for extended duration of phases away from the steady state.

\subsection{Robustness}
Furthermore we are interested how robust the presented adaptive simplicial Snowdrift model is
against the introduction of irrational actions of players.
Such irrational actions are characterized by a non-increasing payoff for the selected player.
For example, rewiring to defecting $\defect$-nodes leads to the same or smaller payoff.
Here we consider only cases, where the \emph{best response simplex}-action is replaced with 
one of the following strategies with some probability $\alpha \in [0, 1]$.
In particular,
for $\alpha = 0$ the original adaptive simplicial Snowdrift model is obtained and
for $\alpha = 1$ all best response actions on simplices are replaced.

For this we consider
(i) the all-defecting strategy which corresponds to the best response
strategy in the Prisoners Dilemma,
(ii) the worst response strategy, where the opposite of best response in simplices is considered,
(iii) the all-cooperating strategy, 
(iv) the adaption to the majority action within the simplex,
(v) copying the neighbors action,
(vi) always changing the action,
(vii)--(ix) increasing the number of $00$-, $01$-, or $11$-edges, and
(x)--(xii) decreasing the number of $00$-, $01$-, or $11$-edges, respectively.

Each of these extensions leads to an additional terms in the moment equations which are proportional to
$\alpha$ and depend on the considered action (i)--(xii). We consider the resulting closed moment equations
using the approximations of Appendix~\ref{app:moment_approximations}.
The numerical bifurcation analysis of these extended models is performed in MatCont, see
Appendix~\ref{app:matcont}, for parameters $N \in [3, 10^6]$, $M = \mu N$ with $\mu \in (1, N(N-1))$, 
$\phi \in (0, 1)$, $\rho \in (0, 1)$, $\sigma \in (0, 1)$, and $\alpha \in (0, 1)$.
The results (not shown)
suggest that the non-trivial steady state remains stable for all reasonable parameter choices
and is only shifted under variation of $\alpha$. Yet, the same transient excursions due to weakly
stable eigenvalues can be possible as discussed above.
Hence we conclude that the dynamics of the adaptive simplicial Snowdrift model is stable,
even if the multiplayer interactions are perturbed.
%


\section{Summary and outlook}
\label{sec:summary}

In summary, we have proposed a multiplayer Snowdrift model on adaptive simplicial complexes. This model contains rewiring rules, changing the topology, and best response rules, changing the state of the system. For the lowest moments, the numbers of cooperators $\mcoop$, the number of edges between cooperators $\mcc$ and the number of edges between defectors $\mdd$ we have derived the moment equations based on the dynamical rules of adaption. Based on the structure of simplicial complexes, we have derived closure relations with pairwise approximation, i.e., relying on the numbers of labeled nodes
(of type $0$ or $1$) and the number of labeled edges (of type $00$, $01$, and $11$). We have confirmed by numerical comparison that the corresponding system of equations leads to a good approximation of
the system, even though only the lowest moments are considered. Since the closed system is analytically not fully solvable, we have shown for specific limiting subsystems, where only one of the considered actions takes place, that the non-trivial steady state point is stable for reasonable hypergraphs as long as (sufficient) rationality governs the dynamics. These results have been further supported with numerical stability analysis of the full system including direct numerical simulations of the full hypergraph dynamics as well as numerical continuation calculations for the reduced moment equations. In summary, we have observed a quite surprising robustness of non-trivial steady states of adaptive multiplayer games, which we attribute to sufficient rational behavior for each player. I.e., even some higher-order interaction rules with irrationality were not able to de-stabilize the non-trivial steady state. 

This suggests two interesting conclusions: (a) Complex socio-economic systems with many players and higher-order interactions might be quite robust, at least if each player/agent has access to enough information to make sufficiently rational decisions and perturbations from the steady states are sufficiently small, and (b) although several examples of de-stabilization of network dynamical systems by higher-order interactions have been found, one should not overlook the possibility that the stability of higher-order systems could often reduce to more classical systems with pairwise interactions. Both conclusions are somewhat re-assuring as there are (a) indeed long periods of socio-economic stability in data, and (b) we can still hope to rely on results for more classical pairwise models for certain modeling situations. 

Of course, we emphasize that our study is only a step towards a deeper understanding of adaptivity in models with higher-order interactions. Many open questions remain, and we mention a few natural ones. From a structural perspective, it would be important to generalize our results from simplicial complexes to more general hypergraphs. A similar remark applies to the derivation of more detailed sequences of moment equations going beyond second-order moments. Unfortunately, both generalizations become quite involved as the combinatorial complexity increases quickly, so developing an efficient mathematical scheme to algorithmically generate moment equations for hypergraphs would be very desirable. From the viewpoint of applications, one could investigate other multiplayer games. In fact, we have also cross-validated our results in variants of the Prisoner's Dilemma~\cite{MADS} but the results/conclusions are similar, so we have not presented them here. Yet, there could be many other multiplayer games arising in various fields, where adaptivity might have a different impact. In this regard, it would also be interesting to develop a more precise quantifier, how adaptivity and the amount of rational agent behavior influence stability or can induce bifurcations, i.e., trying to use the modeling framework we have started to investigate crucial modeling choices in behavioral socio-economic systems.

\acknowledgements

KC and CK thank the VolkswagenStiftung for support via the grant ``Self-Dynamics of Self-Adapting Networks'' within a Lichtenberg Professorship awarded to CK. 

\appendix 

\section{Derivation of moment equations}
\label{app:derivation-moments}
In this section we derive moment equations for the adaptive simplicial Snowdrift model.
We are particularly interested in the lowest moments, i.e., the expected numbers
$\mcoop$ and $\mdefect$ of $\coop$ and $\defect$-nodes, as well as
the expected numbers of edges $\mcc$, $\mcd$, and $\mdd$.
Recall that we consider a fixed total number of nodes
$N = \mcoop + \mdefect$  and
a fixed number of edges $M = \mcc + 2\mcd + \mdd$.
Therefore it is sufficient to derive moment equations for $\mcoop$, $\mcc$
and $\mdd$.

\subsection{Evolution of $\mcoop$}
The number of cooperating nodes $\mcoop$ changes due to best response on edges and
best response on simplices.
Let $\Pxyany$ be the probability that a randomly chosen edge $(i,j) \in \edges$ of type $xy$, i.e.,
$X(i) = x, X(j) = y$, is
part of at least one simplex and
$\Pxyno$ be the probability that it is not part of any simplex. 

\paragraph{Best response on edges}
The number of $\coop$ nodes increases by one
for best response on $11$-edges, and decreases by one
for best response on $00$-edges.
Based on the proposed dynamics
the probability to take an action on edges is given by the probability
to be not in any simplex $\Pno{xy}$ plus the probability 
to be in any simplex $\Pany{xy}$ times the probability of not taking
a simplex action $(1-\rho)$, which gives
$\Pno{xy}
+ (1-\rho)\Pany{xy} = 1 - \rho\Pany{xy}$ where $\Pno{xy} + \Pany{xy} = 1$ has been used.
This leads to the following
contributions to $\frac{\txtd}{\txtd t} \mcoop$,
\begin{equation}
    \begin{split} 
        a^-_0 &= \frac{\mcc}{M} \left(1-\rho\Pany{00}
        \right)(1 - \phi),
        \\
        a^+_0 &= \frac{\mdd}{M} \left(1 - \rho\Pany{11}
        \right)(1 - \phi),
    \end{split} 
\end{equation}
where $\mcc/M$ and $\mdd/M$ are the probabilities that
the chosen edge is of type $00$ and $11$, respectively.
The other factors account for the probability to perform
best response on edges.

\paragraph{Best response on simplices}
For best response in simplices we define
$\Psimp{xy}{d}$
the probability
that an edge of type $xy$ is part of a
simplex with $d$ defecting nodes, conditioned that it is
part of at least one simplex.
Best response on simplices decreases the number of cooperating
nodes, if the selected node is in state $\coop$ and at least
one other node is cooperating. Similarly, it increases
the number of $\coop$ nodes, if all nodes are defecting.
This gives the following contributions to $\frac{\txtd}{\txtd t} \mcoop$,
\begin{equation}
    \begin{split} 
        b^-_0 &= \frac{\mcc}{M} \Pany{00}\rho 
        \left(\Psimp{00}{0} + \Psimp{00}{1}\right)
        + \frac{\mcd}{M} \Pany{01} \rho \Psimp{01}{1},
        \\
        b^+_0 &= \frac{\mdd}{M} \Pany{11} \rho \Psimp{11}{3}.
    \end{split}
\end{equation}
Note that $\mcd / M$ is the probability to select an edge
which goes from a $0$-type edge to a $1$-type edge.
Since we consider undirected graphs both type of edges,
$01$ and $10$, exist in $\edges_2$
for each such pair of connected nodes.
Since we perform best response on the first node
of the selected edge, it is necessary to distinguish
$01$ from $10$ edges. However, since $\mcd = \mdc$ the probabilities to select either are the same.

\paragraph{Resulting equation for $\mcoop$}

Altogether, the evolution of the number of cooperating
$0$-nodes in the Adaptive Simplex Snowdrift Model is governed by the differential equation
$\frac{\txtd}{\txtd t} \mcoop = - a^-_0 + a^+_0  - b^-_0 + b^+_0$,
which is equivalent to Eq.~\eqref{equ_SD_0}.

\subsection{Evolution of $\mcc$}
The number of $00$-edges $\mcc$ changes directly due to
rewiring, as well as directly and
indirectly due to best response on edges and simplices.

\paragraph{Rewiring}
The number of $00$-edges increases, if a $01$-edge is
rewired to a $00$-edge. We do not consider the opposite
direction, since this would decrease the payoff.
Respecting the probability to choose rewiring
of edges (either in or outside of any simplex),
we obtain the following contribution to
$\frac{\txtd}{\txtd t} \mcc$,
\begin{equation}
    a_{00}^+ = \frac{\mcd}{M}\left(1 - \rho\Pany{01} \right) \phi,
\end{equation}
where again it is respected that only $01$-edges rewire to $00$ edges, but not $10$-edges, since in the latter case the first node which performs the action is already connected to a $0$-type node.
Note for each rewired edge the number of both $01$ and $10$-edges decreases by one
    and the number of $00$-edges increases by $2$, such that the total contribution
    is $2a_{00}^+$.

\paragraph{Best response on edges}
The change in the number of cooperating nodes $\mcoop$
also influences the number of $00$-edges.
In particular, if a cooperating node changes from
$\coop$ to $\defect$, all its adjacent $00$-edges to
other $\coop$-nodes become $01$-edges.
The contribution to $\mcc$ from the switching node
comes from its other neighbors of type $0$.
For this, let $\excessn{x}{y}{z}{n}$ be the expected number of
$z$-type neighbors for a node $i$ with type $X(i) = x$, which is already connected to $n$ nodes $j$ with type $X(j) = y$.
Together with the probability of best response on edges,
this gives the total number of added or removed
$\mcc$ connections.
Altogether, this leads to
\begin{equation}
    b_{00}^- 
    = - a_0^-\left(1 + \excessn{0}{0}{0}{1}\right), \quad
    b_{00}^+ 
    = a_0^+ \excessn{1}{1}{0}{1},
\end{equation}
accounting for both direct and indirect contributions.

\paragraph{Best response in simplices}

Best response in simplices also changes the number of
$00$-edges directly and indirectly.
If the simplex is $\simplex_0$, two $00$-edges are destroyed
due to best response.
For $\simplex_1$ simplices changing the node state from
$0$ to $1$ destroys one $00$-edge in each case.
Note that within the simplex best response does not create new $00$-edges.
This implies the contribution
\begin{equation}
    c_{00}^- =
    \frac{\mcc}{M} \Pany{00}\rho 
    \left(2\Psimp{00}{0} + \Psimp{00}{1}\right)
    + \frac{\mcd}{M} \Pany{01} \rho \Psimp{01}{1}.
\end{equation}

The indirect influence comes from the changes of edge types
outside of the simplex due to the change of node state
within the simplex.
For this the simplex excess degree
$\simplexcess{x}{n}{y}$
is defined \cite{MAAB} as
the expected number of additional neighbors $j$ with type $X(j) = y$ of a node with type $X(i) = x$, given that it is part of a simplex of type $n$.
In particular, the $x$-node is within the simplex already
connected to $n - \delta_{x1}$ neighbors of type $1$,
where  $\delta_{ij}$ is the Kronecker delta,
and $3 - n + \delta_{x1}$ neighbors of type $0$.

With this, the indirect contributions to the change
of $00$-type edges can easily be written in terms of
the probabilities for best response in the specific
simplices multiplied with the expected number of
additional $0$-type neighbors.
We end with
\begin{align}
    d_{00}^- &=  \phantom{+} \frac{\mcc}{M} \Pany{00} \rho \
    \left[\Psimp{00}{0}\simplexcess{0}{0}{0}
    + \Psimp{00}{1}\simplexcess{0}{1}{0}
    \right] \\
    &\phantom{=} + \frac{\mcd}{M} \Pany{01} \rho\ 
    \Psimp{01}{1}\simplexcess{0}{1}{0},
    \\
    d_{00}^+ &=  \phantom{+} \frac{\mdd}{M} \Pany{11} \rho \ \Psimp{11}{3}
    \simplexcess{1}{3}{0},
\end{align}

\paragraph{Resulting equation for $\mcc$}
Altogether, the evolution of the number of 
$00$-nodes in the Adaptive Simplex Snowdrift Model is governed by the differential equation
$\frac{\txtd}{\txtd t} \mcc = 2\left(a^+_{00} - b^-_{00} + b^+_{00} - c^-_{00}
- d^{-}_{00} + d^{+}_{00}\right)$,
where the factor of $2$ reflects the double counting of undirected edges.
Altogether, the change in the number of $00$-edges in the adaptive simplicial Snowdrift model
is given by Eq.~\eqref{equ_SD_00}.

\subsection{Evolution of $\mdd$}
Similar as for $\mcc$, the number of $11$-edges changes
due to rewiring and due to best response on edges and
simplices, directly and indirectly.
Therefore the derivation follows analogously.
Summing up direct and indirect contributions leads to
the differential equation in Eq.\eqref{equ_SD_11}.

We emphasize that these three equations for the lowest
moments depend on higher order moments, such as
the distribution of triples through the excess degrees,
and the distribution of simplices of different types.

\section{Moment closure approximations}
\label{app:moment_approximations}

In the following the moment equations are closed at the level
of triples, such that they right-hand side of the ODEs
depends on the lowest moments
$\mcoop$, $\mdefect$, $\mcc$, $\mcd$, $\mdd$, only.
This means that the full information on the network structure
and simplex distribution is reduced to information on the
pairs~\cite{grossAdaptiveNetworks2009,%
    kuehnMomentClosureBrief2016,%
    kissMathematicsEpidemicsNetworks2017,demirelMomentclosureApproximationsDiscrete2014}.
Note that we consider a system with a fixed number of
nodes $N = \mcoop + \mdefect$, a fixed number of
edges $M = \mcc + 2\mcd + \mdd$, and also a fixed number of
simplices $S$.

\subsection{Approximation of excess degrees}

\paragraph{Approximation of $\mathcal{E}^{(a,b)}_1(c)$}
By definition, the excess degree $\excessn{a}{b}{c}{1}$
is the expected number of $c$-type  neighbors of an
$a$-type node which already has one $b$-type neighbour.
Thus, it is equivalent to the expected number of $cab$-triplets containing the $ab$-edge.
Let $n_{abc}$ be the number of triplets with states $X(i)=a$, $X(j)=b$, and $X(k)=c$,
    i.e.,
    $n_{abc} = \sum_{i,j,k = 1}^N A_{kj}A_{ji}\delta_{X(i),a}\delta_{X(j),b}\delta_{X(k),c}$.
    Note that symmetry implies $n_{001} = n_{100}$ and $n_{011} = n_{110}$.
    The expected number of $cab$-triplets $n_{cab}$ is approximately given
    by the number of $ab$-edges multiplied with the expected number of $c$-neighbors
    of $a$. Therefore we have
\begin{equation}  
    \excessn{a}{b}{c}{1} \approx
    \frac{n_{cab}}{n_{ab}}.
\end{equation}

\begin{figure}[t!]
    \begin{overpic}[scale=1]
        {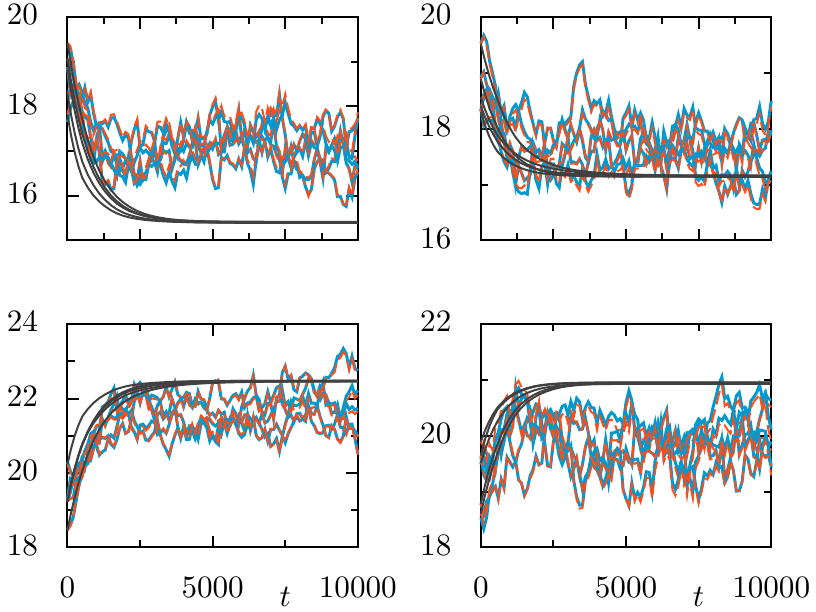}
    \put(15, 76) {(a) $\excessn{0}{0}{0}{1}$}  

    \put(63, 76) {(b) $\excessn{1}{1}{0}{1}$}  

    \put(15, 38) {(c) $\excessn{0}{0}{1}{1}$}  
    
    \put(63, 38) {(d) $\excessn{1}{1}{1}{1}$}  
    \end{overpic}

    \caption{Approximation of excess degrees $\excessn{a}{b}{c}{j}$ for same simulation as in Fig.~\ref{fig:fig1}.
    Shown are values directly obtained from simulation
    (blue lines), compared to
    the approximations in Eq.~\eqref{eq:approx_excess}
    using $n_{x}$ and $n_{xy}$ from simulation (orange dashed lines) and from the closed moment equations (black lines).
        The agreement of the approximation in
        Eq.~\eqref{eq:approx_excess} with the simulation is very good,
        in particular if the current values of the lowest moments
        are used.
        Differences to the closed moment result are of similar kind
        as in Fig.~\ref{fig:fig1}.
    Interestingly we find that  $\excessn{0}{0}{1}{0} \approx \excessn{1}{1}{1}{0} - 1$
    and $\excessn{0}{0}{1}{1} \approx \excessn{1}{1}{1}{1} + 1$.
    This means that the expected number of $0$-type neighbors is similar for $0$-type
    and $1$-type nodes, and similar for the number of $1$-type neighbors.
    The proposed closure is therefore only justified for large average degrees $\degavg$,
    when this difference of $\pm 1$ becomes negligible.}
    \label{fig:fig_Excess1}
\end{figure}
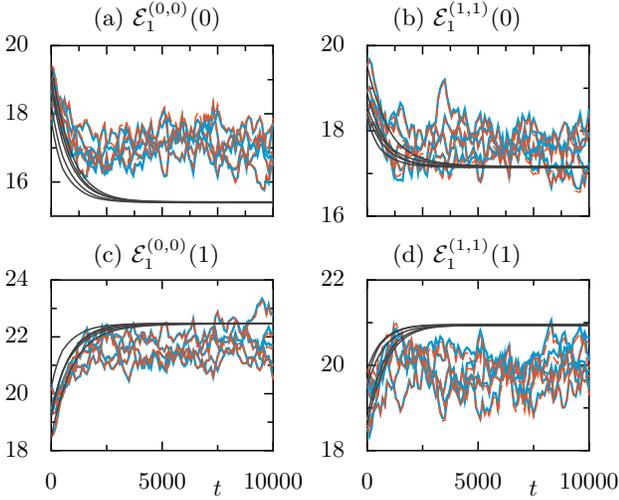

    Furthermore, the numbers of triplets $n_{cab}$ are approximated in
    terms of pairs and nodes as follows.
    The proportion of edges from $a$-type to $b$-type nodes is
    given by the number of $ab$-edges divided by the total number
    of edges starting at $a$, which is approximately $\degavg n_{a}$.
    Thus we have $\frac{n_{ab}}{\degavg n_a}$.
    The probability for two neighbors of some $a$ type node to be of type
    $c$ and $b$, respectively, is thus approximately
    $\frac{n_{ca}n_{ab}}{\degavg(\degavg - 1) n_a^2}$.
    For the total number of $cab$-triplets we have to multiply this with the
    possible number of ways to choose the neighbors, which is $\degavg(\degavg - 1)$,
    and with the total number of $a$-nodes, and we get
\begin{equation}
    n_{cab} \approx \frac{n_{ca} n_{ab}} {n_a}
\end{equation}
such that
\begin{equation}
    \begin{split}
        &\mathcal{E}^{(0,0)}_1(0) \approx \frac{\mcc}{\mcoop},~
        \mathcal{E}^{(1,1)}_1(0) \approx \frac{\mcd}{\mdefect},\\
        &\mathcal{E}^{(0,0)}_1(1) \approx \frac{\mcd}{\mcoop},~
        \mathcal{E}^{(1,1)}_1(1) \approx \frac{\mdd}{\mdefect}.
        \label{eq:approx_excess}
    \end{split}
\end{equation}

We emphasize that for this pair approximation of the triplets the approximative excess degrees 
of $a$-type nodes are independent on the number of already connected $b$-nodes
and therefore are simply given by the expected number of $c$-node neighbours.

Note that $\mcoop=0$ also implies $\mcc = 0$ and $\mcd = 0$ and
    the terms containing the excess degrees do not contribute.
For illustration of the agreement of the closure with simulation
see Fig.~\ref{fig:fig_Excess1}.

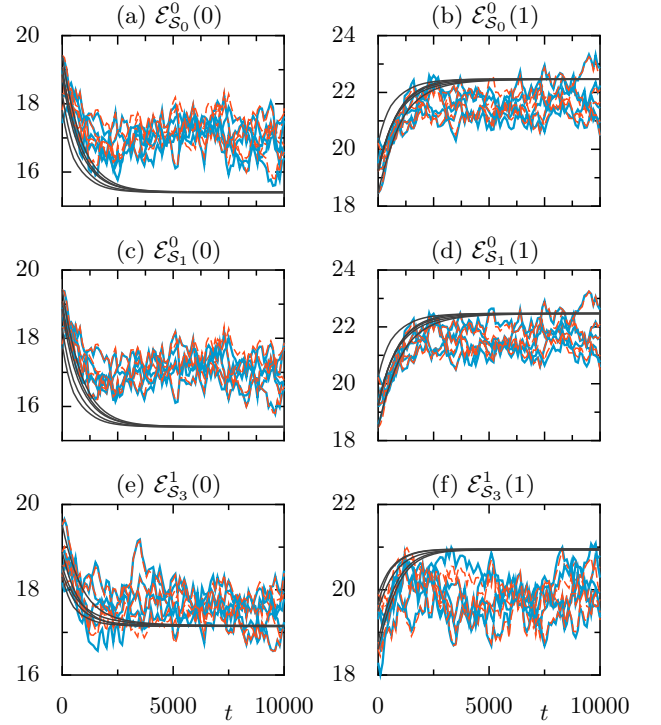
\begin{figure}[t!]
    \begin{overpic}[scale=1]
        {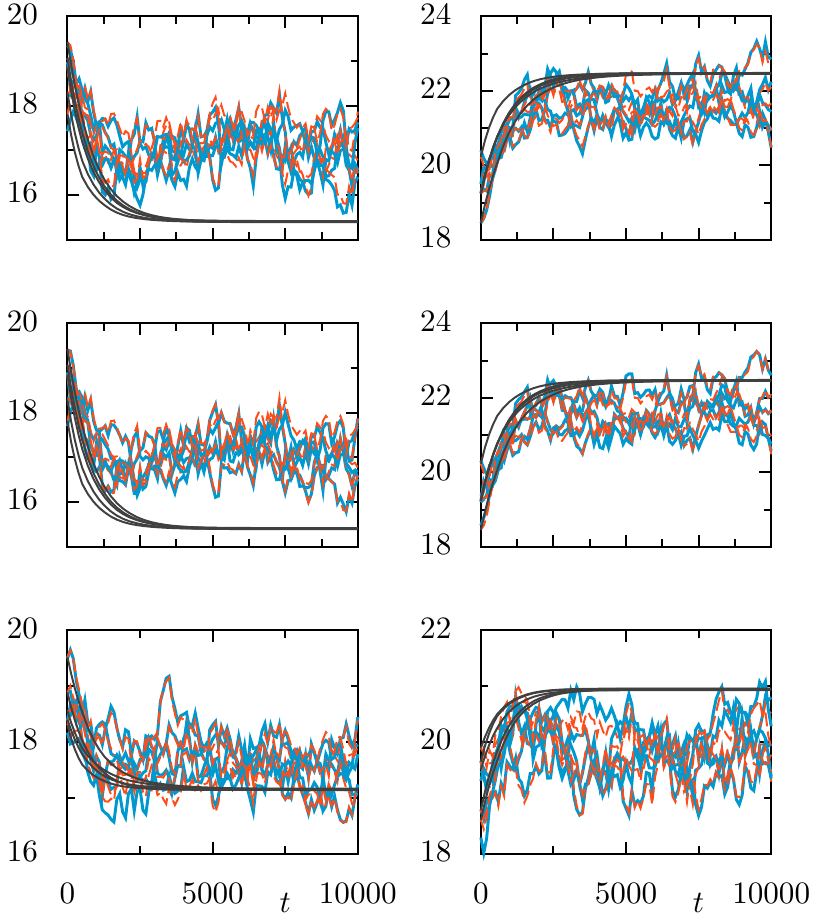}
    \put(15, 100) {(a) $\simplexcess{0}{0}{0}$}  
    \put(60, 100) {(b) $\simplexcess{0}{0}{1}$} 

    \put(15, 66.5) {(c) $\simplexcess{0}{1}{0}$}  
    \put(60, 66.5) {(d) $\simplexcess{0}{1}{1}$}  

    \put(15, 33) {(e) $\simplexcess{1}{3}{0}$}  
    \put(60, 33) {(f) $\simplexcess{1}{3}{1}$}  
    \end{overpic}
    \caption{Approximation of simplex excess degrees
        $\simplexcess{x}{j}{y}$
        for same simulation as in Fig.~\ref{fig:fig1}.
        Shown are values directly obtained from simulation
        (blue lines), compared to
        the approximations in Eq.~\eqref{eq:approx_excess}
        using $n_{x}$ and $n_{xy}$ from simulation (orange dashed lines) and from the closed moment equations (black lines).
    }
    \label{fig:fig_Excess2}
\end{figure}
\paragraph{Approximation of $\simplexcess{a}{i}{c}$}

The simplex excess degree $\simplexcess{a}{i}{c}$ is the expected number of $c$-type-neighbours of an $a$-node within
a simplex of type $i$. 
In order to apply pair approximation we notice that
an $a$-type node within a simplex of type $i$ is
already connected to $i - \delta_{a,1}$ neighbors of type
$1$.
For example, if $a=0$, $i=0$, we
have $\simplexcess{0}{0}{0} \approx \excessn{0}{0}{0}{2}$
and $\simplexcess{0}{0}{1} \approx \excessn{0}{0}{1}{2}$.
Similarly $\simplexcess{1}{3}{0} \approx \excessn{1}{1}{0}{2}$
and $\simplexcess{1}{3}{1} \approx \excessn{1}{1}{1}{2}$.
If the average degree $\degavg$ in the network is large
these excess degrees are approximately
$\excessn{a}{b}{c}{2} \approx \excessn{a}{b}{c}{1}$.
If $a=0$ and $i=1$, the $a$-node is already connected to
one neighbor of type $0$ and one neighbor of type $1$.
Therefore, we can approximate
$\simplexcess{0}{1}{0} \approx \excessn{0}{0}{0}{1}$,
but also $\simplexcess{0}{1}{0} \approx \excessn{0}{1}{0}{1}$.
In pair-approximation both quantities are the same, 
Similarly we obtain  $\simplexcess{0}{1}{1} \approx \excessn{0}{0}{1}{1}\approx \excessn{0}{1}{1}{1}$.

Thus, for large node degrees $\degavg$ the approximations in Eqs.~\eqref{eq:approx_excess}
from the previous subsection are applied.
We illustrate the time dependence of the simplex-excess degrees
in Fig.~\ref{fig:fig_Excess2},
comparing values from the simulation (colored lines) to the
corresponding results from the moment-closure system (black lines).

\subsection{Approximation of probabilities}
\label{app:approx_prob}

\paragraph{Approximation of $\Pany{xy}$}
First we consider the probability that a $01$-edge is
not part of any simplex.
Intuitively, only simplices in $\simplex_1$ and $\simplex_2$ contain $01$-edges, and there are for each type exactly two
such edges.
Thus, a randomly drawn $01$-edge $e_{01}\in\edges$ is part of 
a specific simplex
$s^\ast \in \simplex_1 \cup \simplex_2$ with probability
\begin{equation}
    \prob(e_{01} \in s^\ast)=\frac{2}{\mcd},
\end{equation}
because $s^\ast$ contains two out of all $\mcd$ edges of type $01$.
Furthermore we note that if the $01$-edge is part of $s^\ast = (i, j, k) \in \simplex$,
then it is also part of any permutative simplex, e.g.,
$\tilde{s} = (j, i, k) \in \simplex$, which is of the same type.
This means that each $3! = 6$ simplices are equivalent.
Let $\tilde{\simplex}_i := \simplex_i|_{\simeq}$ be the set of all unique simplices
defined by this equivalence condition, i.e., $|\tilde{\simplex}_i| = |{\simplex}_i| / 6$.
The probability that the $01$-edge is not part of any simplex 
is thus given by
\begin{align}
    \Pno{01} &= \prob(e_{01} \notin s^\ast \, \forall s^\ast \in \simplex_1 \cup \simplex_2)\\
    &\approx \prod_{s^\ast \in \tilde{\simplex}_1 \cup \tilde{\simplex}_2}
    \prob(e_{01} \notin s^\ast)\\
    &= \left(1 - \frac{2}{\mcd}\right)^{(|\simplex_1| + |\simplex_2|) / 6},
\end{align}
where the approximation assumes that the simplices in $\tilde{\simplex}_i$ are independent.
Secondly, edges of $00$-type can only be part of simplices in $\simplex_0$ and $\simplex_1$.
Recall that $s \in \simplex_0$ contains three $0$-nodes and
three 
$00$-edges {counted twice}, while $s\in \simplex_1$ contains exactly
one $00$ edge {counted twice}.
Thus, a randomly drawn $00$-edge $e_{00} \in \edges$
is part of a simplex $s^\ast_0 \in \simplex_0$ with probability
\begin{equation}
    \prob(e_{00} \in s_0^\ast) = \frac{6}{\mcc}
\end{equation}
and is part of a simplex  $s^\ast_1 \in \simplex_1$
with probability
\begin{equation}
    \prob(e_{00} \in s_1^\ast) = \frac{2}{\mcc}.
\end{equation}
The probability that $e_{00}$ is not part of
any simplex is thus %
\begin{align*}
    \Pno{00} &= \prob(e_{00} \notin s_0\  \forall s_0 \in \simplex_0 ~\wedge ~  e_{00} \notin s_1  \ \forall s_1 \in \simplex_1)\\
    &\approx
    \prod_{s_0\in\tilde{\simplex}_0} \prob(e_{00} \notin s_0)
    \cdot 
    \prod_{s_1\in\tilde{\simplex}_1} \prob(e_{00} \notin s_1)\\
    &= \bigg(1-\frac{{6}}{\mcc}\bigg)^{| \simplex_0| /6} \bigg(1-\frac{{2}}{\mcc}\bigg)^{| \simplex_1| / 6}.
\end{align*} 
Analogously one derives the approximations
\begin{equation}
    \Pno{11} \approx\bigg(1-\frac{{2}}{\mdd}\bigg)^{| \simplex_2| / 6} \bigg(1-\frac{{6}}{\mdd}\bigg)^{| \simplex_3| / 6}.
\end{equation}

Altogether, this approximates the probabilities
of some edge $e_{xy}$ of type $xy$ to be part of any simplex,
\begin{equation}
    \Pany{xy} = 1 - \Pno{xy},
    \label{eq:pany_approx}
\end{equation}
with approximations for $\Pno{xy}$ as given above.
Suitable approximations for the numbers of simplices $|\simplex_i|$ are given in
the end of this section.

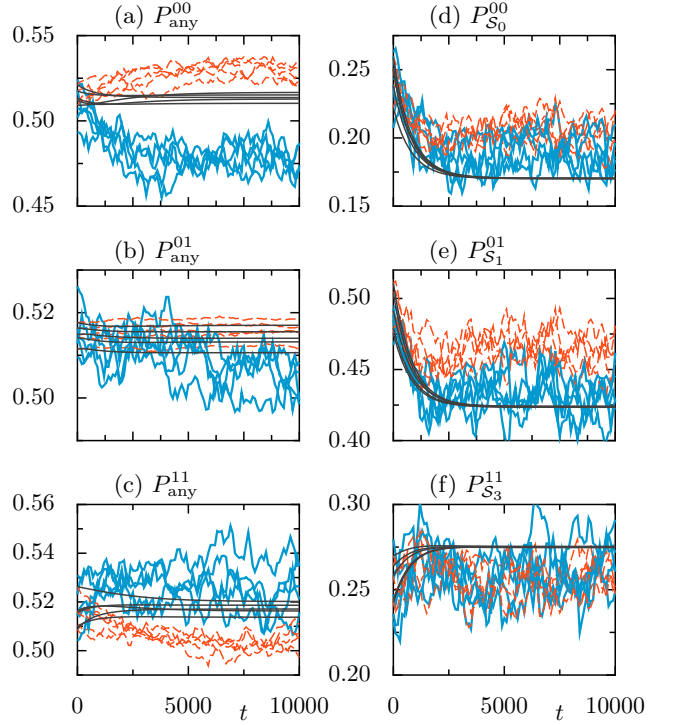
\begin{figure}[t!]
    \begin{overpic}[scale=1]
        {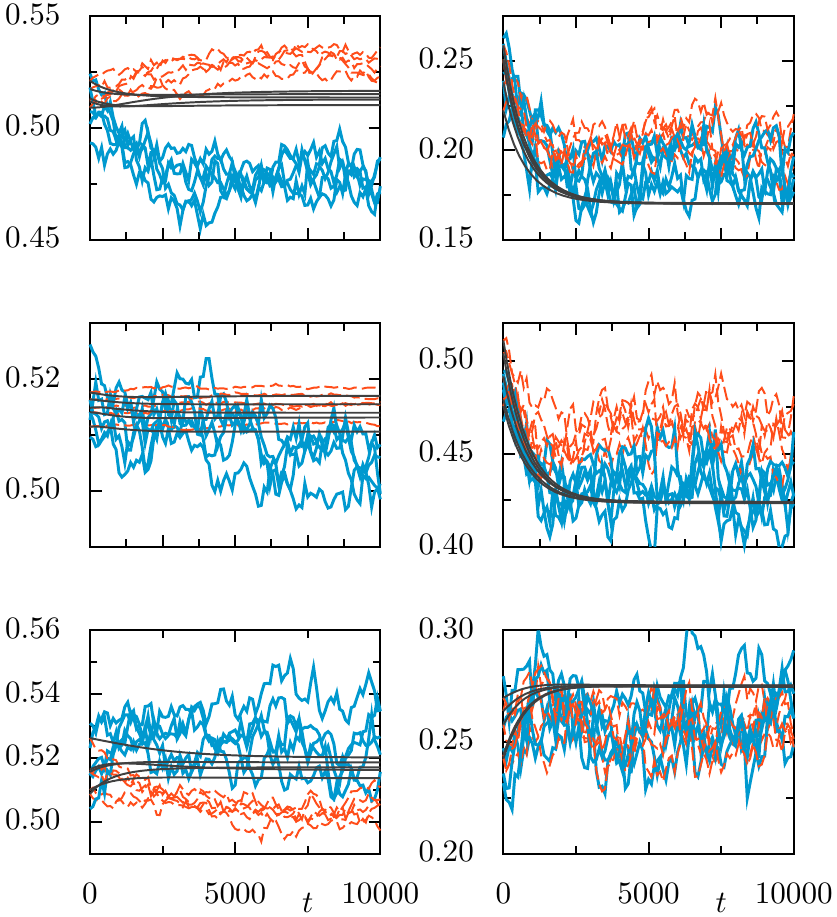}
        \put(15, 100) {(a) $\Pany{00}$}  
        \put(60, 100) {(d) $\Psimp{00}{0}$} 
        
        \put(15, 66.5) {(b) $\Pany{01}$}  
        \put(60, 66.5) {(e) $\Psimp{01}{1}$}  
        
        \put(15, 33) {(c) $\Pany{11}$}  
        \put(60, 33) {(f) $\Psimp{11}{3}$} 
    \end{overpic}
    \caption{Approximation of probabilities for same simulation as in Fig.~\ref{fig:fig1}.
Shown are values directly obtained from simulation
(blue lines), compared to
the approximations for $\Pany{xy}$ in
Eq.~\eqref{eq:pany_approx}
and for $\Psimp{xy}{i}$ in Eq.~\eqref{eq:pany_approx}
using $n_{x}$ and $n_{xy}$ from simulation (orange dashed lines) and from the closed moment equations (black lines).
\linebreak
The most significant deviations between approximation and numerical values
are observed for $\Pany{00}$, for which the approximation is considerably
larger. This is reasonable to expect, since new $00$-edges are created due
to rewiring of $01$-edges, without simultaneously increasing the number
of adjacent simplices. For the corresponding numbers of simplices see Fig.~\ref{fig:fig_NumSimplex}. 
    }
    \label{fig:fig_Probs}
\end{figure}

For a comparison of these approximative proabilities for the three edge types
with the numerical simulation see
Fig.~\ref{fig:fig_Probs}(a)--(c).

\paragraph{Approximation of $\Psimp{xy}{i}$}
\label{app:number_simplices}

Recall that $\Psimp{xy}{i}$ is the probability that an $xy$-type edge is in a specific type $i\in \{0,1,2,3\}$ of simplex, conditioned on the event that the edge is part of at least one simplex.
Considering the possible type of simplices which a specific edge can be part of, these
probabilities are given by
\begin{equation}
    \begin{split}
        \Psimp{00}{0}
        &= \frac{|\simplex_0|}{|\simplex_0| + |\simplex_1|}, \quad
        \Psimp{00}{1}
        = \frac{|\simplex_1|}{|\simplex_0| + |\simplex_1|}\\
        \Psimp{01}{1}
        &= \frac{|\simplex_1|}{|\simplex_1| + |\simplex_2|}, \quad
        \Psimp{01}{2}
        = \frac{|\simplex_2|}{|\simplex_1| + |\simplex_2|}\\
        \Psimp{11}{2}
        &= \frac{|\simplex_2|}{|\simplex_2| + |\simplex_3|}, \quad
        \Psimp{11}{3}
        = \frac{|\simplex_3|}{|\simplex_2| + |\simplex_3|}.
    \end{split}
\end{equation}
These approximated probabilities are compared to the values from
numerical simulation in
Fig.~\ref{fig:fig_Probs}(d)--(f), where the following approximations
for the numbers of simplices are considered.

\paragraph{Approximation of $| \simplex_i |$}

For a simplicial complex as considered in this paper, it is possible to
approximate the number of simplices with the numbers of labeled
edges in $\edges_2$.
We assume that the simplices are distributed proportional
to the triangles in the network,
\begin{equation} \label{equ_Si}
    | \simplex_i |  \approx S \cdot \prob(t \in \mathcal{T} _i),~ i\in \{0,1,2,3\}
\end{equation}
where $S$ is the number of simplices, $t$ some triangle and $\mathcal{T}_i$ the set of all triangles of type $i$ in the network.
The number of simplices $S$ is constant in our model.
If a simplex is destroyed due to rewiring, a new simplex is added on the set of triangles.
We want to express the probabilities that some triangle $t$ is of type $i \in \{0,1,2,3\}$,
\begin{equation}
    P(t \in \mathcal{T}_i ) = \frac{|\mathcal{T}_i|}{T},
\end{equation}
in terms of the nodes and edges, where $T =  \sum_{i} |\mathcal{T}_i|$.

For this we consider a triangle $t_{abc}$ with three nodes of status $a, b, c$
as an $abc$-subgraph, where the $a$ and $c$ node are connected.
For example,
the number of $000$-triangles is approximated as the number
$n_{000}$ of $000$-triples, multiplied with the probability that the outer $0$-type nodes
are connected, which approximately is $\mcc / \mcoop^2$.
    This approximation becomes better for large average node degrees $\degavg$.
    Altogether, this leads to
\begin{equation}
    |\mathcal{T}_0| \approx \mccc \frac{\mcc}{\mcoop^2} \approx \frac{\mcc^3}{\mcoop^3}.
    \label{eq:approx_triangle_0}
\end{equation}

    For triangles of type $1$ we have to take into account the contributions from
    $001$-triples, $010$-triples and $100$-triples.
    The $001$-triples contribute with $n_{001}$ multiplied with the probability
    that the outer $0$-node and $1$-node are connected, which approximately is
    $\mcd / (\mcoop \mdefect)$ and the same contribution comes from $100$-triples.
    The $010$-triples contribute with $n_{010}$ multiplied
    with the probability that the outer $0$-type nodes
    are connected, which approximately is $\mcc / \mcoop^2$.
    Altogether we have
\begin{equation}
    |\mathcal{T}_1| \approx
    2 n_{001} \frac{n_{01}}{\mcoop\mdefect}
    + n_{010}\frac{n_{00}}{\mcoop^2}
    \approx 3\frac{\mcc\mcd^2}{\mcoop^2\mdefect}.
\end{equation}
A similar argumentation leads to the following approximations
    for the number of type-$2$ and type-$3$ triangles,
\begin{align}
    |\mathcal{T}_2| &\approx
    2 n_{011} \frac{n_{01}}{\mcoop\mdefect}
    + n_{101}\frac{n_{11}}{\mdefect^2}
    \approx 3\frac{\mdd\mcd^2}{\mcoop\mdefect^2},\\
    |\mathcal{T}_3| &\approx
    n_{111} \frac{n_{11}}{\mdefect^2} \approx \frac{\mdd^3}{\mdefect^3}.
    \label{eq:approx_triangle_3}
\end{align}

\begin{figure}[t!]
    \begin{overpic}[scale=1]
        {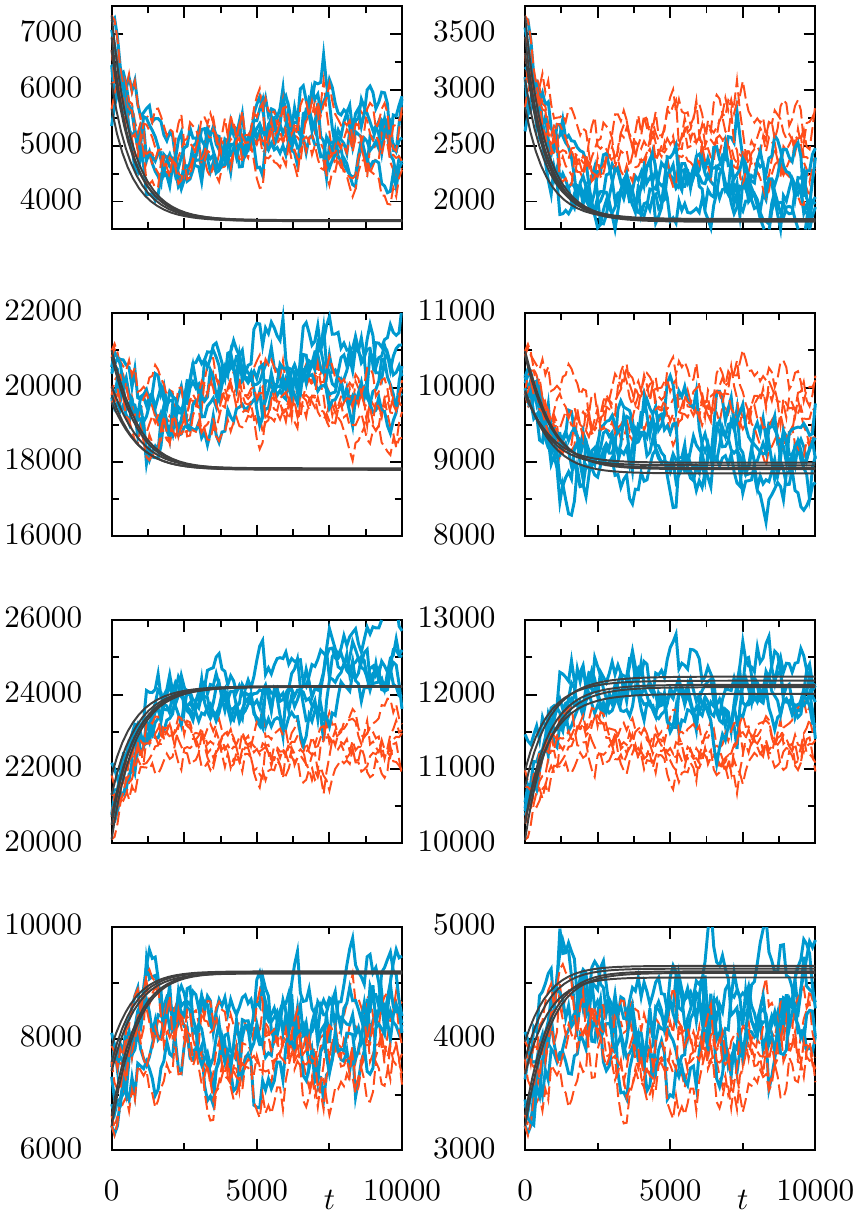}
\put(15, 101) {(a) $|\mathcal{T}_0|$}  
\put(50, 101) {(b) $|\simplex_0|$} 

\put(15, 75.5) {(c) $|\mathcal{T}_1|$}  
\put(50, 75.5) {(d)  $|\simplex_1|$}  

\put(15, 50) {(e) $|\mathcal{T}_2|$}  
\put(50, 50) {(f) $|\simplex_2|$}

\put(15, 24.5) {(e) $|\mathcal{T}_3|$} 
\put(50, 24.5) {(f) $|\simplex_3|$}
    \end{overpic}
    \caption{Approximation of numbers of triangles and 
        simplices of different types
        for same simulation as in Fig.~\ref{fig:fig1}.
        Shown are values directly obtained from simulation
        (blue lines), compared to
        the approximations for $|\mathcal{T}_i|$ in
        Eqs.~\eqref{eq:approx_triangle_0}--\eqref{eq:approx_triangle_3}
        and for $|\simplex_i|$ in Eq.~\eqref{eq:approx_num_simplex}
        using $n_{x}$ and $n_{xy}$ from simulation (orange dashed lines) and from the closed moment equations (black lines). The most significant deviations between approximation and numerical values
    are observed for $|\mathcal{T}_2|$,
    for which the approximation (orange line) is considerably smaller.
This is reasonable to expect, since simplices (and thus triangles)
of type $2$ are not destroyed
by best response in simplices, which is not covered by
uniformity assumptions and pair approximation.
    }
    \label{fig:fig_NumSimplex}
\end{figure}
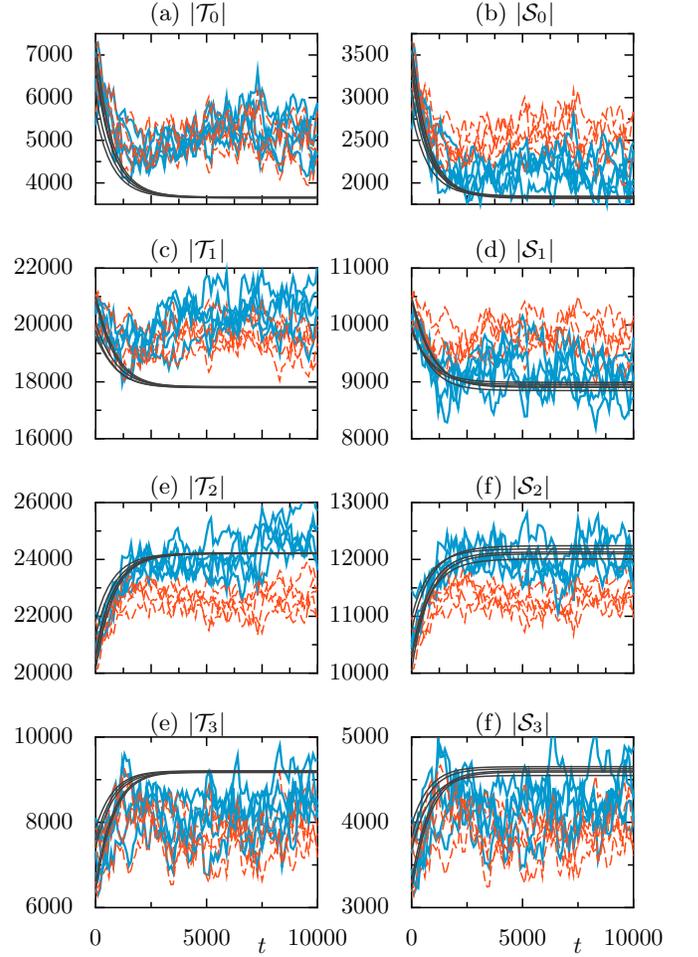
These approximations are inserted into Eq.~\eqref{equ_Si}
and we obtain with
$|\simplex_i| \approx \frac{S|\mathcal{T}_i|}{T}$ and
\begin{align}
    T &\approx
    \frac{\mcc^3}{\mcoop^3} +
    3\frac{\mcc\mcd^2}{\mcoop^2\mdefect} + 3\frac{\mdd\mcd^2}{\mcoop\mdefect^2}
    +    \frac{\mdd^3}{\mdefect^3} \\
    &= \nonumber
    \frac{\mcc^3 \mdefect^3 + 3\mcoop\mdefect^2\mcc\mcd^2
        +3\mcoop^2 \mdefect \mdd\mcd^2 + \mcoop^3\mdd^3
    }{\mcoop^3\mdefect^3}
\end{align}
the following approximations for the different type of simplices,
\begin{align}
    \label{eq:approx_num_simplex}
    \begin{split}
        |\mathcal{S}_0| &\approx
        \frac{S\cdot \mcc^3\mdefect^3}
        {\mcc^3\mdefect^3 + 3\mcc\mcd^2\mcoop\mdefect^2
            + 3\mdd\mcd^2\mcoop^2\mdefect + \mdd^3\mcoop^3},\\
        |\mathcal{S}_1| 
        &\approx  \frac{S \cdot 3\mcc \mcd^2 \mcoop \mdefect^2}
        {\mcc^3\mdefect^3 + 3\mcc\mcd^2\mcoop\mdefect^2
            + 3\mdd\mcd^2\mcoop^2\mdefect + \mdd^3\mcoop^3},\\
        |\mathcal{S}_2|
        &\approx  \frac{S \cdot 3 \mdd \mcd^2 \mcoop^2 \mdefect}
        {\mcc^3\mdefect^3 + 3\mcc\mcd^2\mcoop\mdefect^2
            + 3\mdd\mcd^2\mcoop^2\mdefect + \mdd^3\mcoop^3},\\
        |\mathcal{S}_3| 
        &\approx  \frac{S \cdot \mdd^3\mcoop^3}
        {\mcc^3\mdefect^3 + 3\mcc\mcd^2\mcoop\mdefect^2
            + 3\mdd\mcd^2\mcoop^2\mdefect + \mdd^3\mcoop^3}.
    \end{split}
\end{align} 
Note that for these equations we assume that $n_{0,1}\neq 0$ and $n_{00,01,11}\neq 0$.
Let us emphasize that due to rewiring of edges the
actual number of triangles $T$ and also its approximation
are not constant in time.
However, the considered approximations for the numbers of simplices
satisfy $\sum_{i=0}^3 |\simplex_i| = S$ for all times.
The approximations for the number of triangles and simplices
are illustrated in Fig.~\ref{fig:fig_NumSimplex}. 
For the specific probabilities $\Psimp{xy}{i}$ which are used in the closed moment equations
we get
\begin{equation}
    \label{eq:psimp_approx}
    \begin{split}
        \Psimp{00}{0}
        &\approx  \frac{\mcc^2\mdefect}{\mcc^2\mdefect + 3 \mcd^2 \mcoop},\\
        %
        \Psimp{01}{1}
        &\approx  \frac{\mcc\mdefect}{ \mcc\mdefect + \mdd \mcoop},\\
        \Psimp{11}{3}
        &= \frac{\mdd^2 \mcoop}{3 \mcd^2 \mdefect + \mdd^2 \mcoop}.
    \end{split}
\end{equation}

\section{Stability analysis with MatCont}
\label{app:matcont}
MatCont is a Matlab software project for the numerical continuation and bifurcation study of continuous and discrete parametrized dynamical systems, \cite{dhoogeNewFeaturesSoftware2008}.
In MatCont the adaptive simplicial Snowdrift model is implemented via its closed moment equations in terms of relative variables.
This is done for the system of closed moment equations, Sec.~\ref{sec:moment_closure},
and also separately for
the simplified parameter choices from Sec.~\ref{sec:stability_analysis}.

For the simulations of the closed moment equations initial values are selected on a grid like
from the admissible range or in small neighbourhoods around already known steady states.  
The analysis allows to make statements about the local and also global stability.
Furthermore, for the bifurcation analysis,
we consider initial conditions at a steady state for a specific choice of parameters.
This is followed by parameter variation, where one of the parameters changes while the others remain
constant.
The real parts of the eigenvalues of the Jacobian of the system reveal the stability of the
steady state and MatCont thereby detects possible bifurcations.
This  procedure is done 
for several steady states, choosing values of $\phi$ and $\rho$ on a grid in $[0,1]^2$.
These parameters control the strength of influence of the three different adaptive operations.
We also consider different parameters $N$, $\mu$ with $M=2 \mu N$ and $\sigma$ with $S= \sigma E(T) \in \mathbb{N}_{>0}$ which determine the properties of the underlying network.

\end{document}